\documentclass[11pt,reqno,a4paper]{article}
\usepackage{geometry}                % See geometry.pdf to learn the layout options. There are lots.
\usepackage{graphicx}
\usepackage{subfig}
\usepackage{amssymb}
\usepackage{amsmath}
\usepackage{epstopdf}
\usepackage{verbatim}
\usepackage{hyperref}
\usepackage{authblk}
\usepackage{floatrow}
\usepackage{caption}

\title{Anomaly Induced Inflation in the Minimal Quartic Extension of Einstein's Gravity}

\author{Kerim Demirel\footnote{email: \href{mailto:demirelk@metu.edu.tr}{demirelk@metu.edu.tr}} \,  and Bayram Tekin\footnote{email: \href{mailto:btekin@metu.edu.tr}{btekin@metu.edu.tr}}}
\affil{Department of Physics, Middle East Technical University \\ 06800, Ankara, Turkey }
\date{}                                           % Activate to display a given date or no date

\begin{document}
\maketitle

\begin{abstract}
\noindent A particular extension of Einstein's General Relativity up to and including quartic terms in the curvature tensor is minimal in the sense that it has a unique maximally symmetric vacuum and only a massless spin-2 excitation in its spectrum around the vacuum. We study the inflation phase of the universe in this minimal quartic extension of Einstein's gravity in the presence of trace anomaly terms coming from the Standard Model fields and the fields of the Minimal Supersymmetric Standard Model. We show that the theory allows a quasi-de Sitter phase with sufficient $e$-foldings.
  
\end{abstract}
\tableofcontents

\section{Introduction}

The extreme homogeneity and the isotropy in the Cosmic Microwave Background (CMB) data \cite{wmap} can be understood if the universe had gone through a brief period of accelerated expansion phase, or if the universe were a lot older. The latter option is unlikely. Within General Relativity such an expansion requires the existence of additional scalar fields, such as the "inflaton" that first inflates the universe, gets affected by the inflation, and creates friction to exit the inflation phase gracefully. Another possible path to inflation is to modify the underlying gravity theory. Here we follow this second route and study the inflationary phase of a modified gravity theory built on the powers of the curvature tensors up to and including the quartic terms. The theory that we shall study here was initially obtained as a subset of Born-Infeld-type gravity theory that was judiciously constructed to have all the good features of Einstein's theory, such as the existence of a unique maximally symmetric vacuum and the non-existence of degrees of freedom besides a massless spin-2 graviton, but is much better behaved in the ultraviolet region of gravity. With properly chosen free parameters, Born-Infeld gravity was shown to reduce to a quartic theory in the form of $\mathcal{F}(R,\mathcal{G})$ gravity theories where $\mathcal{G}$ stands for Gauss-Bonnet scalar. This particular quartic theory has a unique de Sitter vacuum, but it does not have an inflationary phase, that is it does not have a quasi-de Sitter solution with enough $e$-foldings. To remedy this, we shall resort to the conformal anomaly that produces additional degrees of freedom besides the massless spin-2 graviton. This scenario was originally proposed by Starobinsky in 1980 \cite{Starobinsky}. We shall adopt it to our theory, in which, inflation is driven by a quantum anomaly arising from the matter fields' renormalization procedure. Renormalization results in an anomalous energy-momentum tensor that could generate an unstable vacuum with a higher energy density so that the Universe makes a transition from this into the true vacuum state by inflating from a microscopic size to a macroscopic size.

The paper is organized as follows. In the next section, we first describe the full Born-Infeld gravity theory and its cosmological solutions; we show how the theory reduces to a quartic extension (that is the action is a specific polynomial of the curvature tensor up to the fourth power of the Riemann curvature tensor and its contractions) of Einstein's theory. In Section \ref{TA_section}, we introduce and summarize the idea of trace anomaly before embedding it into our theory. Then, we derive the field equations and de Sitter solutions in the presence of trace anomaly. The analysis continues with the linear perturbation around the de Sitter solutions to find the stability conditions. To study the inflationary dynamics, we first examine the phase portrait of the dynamical system to see its general behavior, and then solve the system numerically for particular inflationary solutions. Lastly, by proposing an approximate solution, we study the matter perturbations during inflation. 

%%%%%%%%%%%%%%%%%%%%%%%%%%%%%%%%%%%%%%%%%%%%%%%%%%

\section{The minimal extension of Einstein's gravity}
\noindent
The Born-Infeld (BI) type gravity theories are a class of theoretical frameworks in gravitational physics that extend Einstein's General Theory of Relativity (GR). They emerge from the need to address the inherent singularities present in GR, such as those associated with black holes and the Big Bang, by introducing higher-derivative and higher-curvature corrections to the gravitational action. These corrections are inspired by the Born-Infeld electrodynamics (to be historically accurate these theories were first considered by Schrodinger and later by Eddington in gravity), originally formulated to regularize the self-energy of point charges. BI gravity theories have been studied in the context of string theory and modified gravity theories, offering new perspectives on the fundamental nature of gravity and its behavior in extreme conditions.  The general form of the action of the BI-type gravity theories can be given by
\begin{equation}
    I=\frac{1}{2 \kappa_0 \gamma} \int d^4 x \sqrt{-\operatorname{det}\left(g_{\mu \nu}+ \gamma A_{\mu \nu}\right)},
\end{equation}
where $\gamma$ is the Born-Infeld parameter and  $A_{\mu \nu}$ is a second rank tensor constructed from the curvatures. Of course, a proper $A_{\mu \nu}$ must be found that is consistent with all the known tests of gravity and, that, also does not lead to inconsistencies, such as ghosts in the theory. For example, the naive, obvious choice  $A_{\mu \nu} = R_{\mu \nu}$ leads to massive spin-2 ghosts in the spectrum. How ghosts can be avoided in these types of theories is a rather complicated problem, especially about a de-Sitter or anti-de-Sitter background.  But fortunately, this problem has been fully solved which we discuss below. 

Recently, a type of BI gravity model has been proposed \cite{Gullu4d, tahsin3, tahsin1, Tahsintez,esin} with a unique maximally symmetric solution and {\it only} a single massless spin-2 excitation about this vacuum, as a minimal extension of Einstein’s gravity. With a bare cosmological constant $\Lambda_0$, the action of the model is
\begin{equation}
\label{action}
    I=\frac{1}{2 \kappa_0 \gamma} \int d^4 x\left (\sqrt{-\operatorname{det}\left(g_{\mu \nu}+4 \gamma A_{\mu \nu}\right)}-\left(4 \gamma \Lambda_0+1\right) \sqrt{- g}\right),
\end{equation}
where the second-rank tensor $A_{\mu \nu}$  has 3 undetermined constants, $a,b,c $ and is given as 
\begin{align}
    A_{\mu \nu} &= R_{\mu \nu}+c S_{\mu \nu} \nonumber \\
    &+4 \gamma\left(a C_{\mu \rho \nu \sigma} R^{\rho \sigma}+\frac{c+1}{4} R_{\mu \rho} R_\nu^\rho+\left(\frac{c(c+2)}{2}-2-b\right) S_{\mu \rho} S_\nu^\rho\right) \nonumber \\
    &+\gamma g_{\mu \nu}\left(\frac{9}{8} C_{\rho \sigma \lambda \gamma} C^{\rho \sigma \lambda \gamma}-\frac{c}{4} R_{\rho \sigma} R^{\rho \sigma}+b S_{\rho \sigma} S^{\rho \sigma}\right).
\end{align}
Here, $R_{\mu \nu}$ is the Ricci tensor while $S_{\mu \nu}$ is the traceless Ricci tensor, and $C_{\mu \rho \nu \sigma}$ is the Weyl tensor. The features, formulation of this theory, and its extension to generic $n$ dimensions were laid out in detail in \cite{tahsin3}, which relies on the foundation given in \cite{tahsin1}. Note that this theory is in some sense the minimal BI gravity extension of Einstein's theory as one can add more powers of curvature inside the determinant to obtain non-minimal extensions. Note also that as far as these theories are concerned, the $2+1$ dimensional BI gravity is unique, describing a {\it massive} graviton (instead of a massless one) with a remarkably simple Lagrangian of the form ${\mathcal{L}}= \sqrt{-\det\left(g_{\mu \nu}+ \gamma G_{\mu \nu} \right)}$ \cite{BICQG} where $G_{\mu \nu}$ is the Einstein tensor. The success of this particular theory led us to search for similar theories in four and higher dimensions. 
Let us recapitulate some important aspects of the theory (\ref{action}) that arise from the properties of the determinant and are not shared by most higher derivative gravity theories:

\begin{itemize}

\item The theory has only a single massless spin-2 particle in the spectrum about its flat or (anti)-de Sitter vacuum.  

\item  The effective (dimensionless) cosmological constant ($\lambda$) of the maximally symmetric solution, i.e. $R^{\mu \nu}_{\,\sigma \rho} = \frac{\lambda}{3 \gamma} ( \delta^\mu_\sigma \delta^\nu_\rho- \delta^\mu_\rho \delta^\nu_\sigma) $, obeys the quartic equation  
\begin{equation}
    4 \lambda^4 + 4 \lambda^3 - \lambda + \lambda_0=0,
\label{eff_cosmo}
\end{equation}
which has only one viable solution, i.e. the maximally symmetric vacuum is unique.   

\item  The effective Newton's constant is related to the bare one and $\lambda$ via 
\begin{equation} 
    \kappa = \frac{\kappa_0}{\left(1-4 \lambda\right)\left(1+2\lambda\right)^{2}}.
    \label{eff_newton}
\end{equation}

\item When expanded in small curvature  ($\vert \gamma R \vert \ll 1$), the theory reproduces General Relativity as demanded by low energy observations at the first order; and at the quadratic order, it reproduces  GR modified with the Gauss-Bonnet term, which at the classical level is just GR due to the topological nature of the Gauss-Bonnet term in four dimensions. The specific cubic and quartic order theories we shall work with here were studied recently in \cite{esin} in the context of black hole physics.
\end{itemize}

%%%%%%%%%%%%%%%%%%%%%%%%%%%%%%%%%%%%%%%%%%%%%%%%%%

\subsection{Cosmological solutions and the minimal `quartic' theory}

To investigate the cosmological solutions of the theory, let us consider the simple spatially flat Robertson-Walker metric,
\begin{equation}
    ds^2 = - dt^2 + a(t)^2 d\vec{x} \cdot   d\vec{x}.
\label{RW}
\end{equation}
The field equation coming from the action is highly complicated as given in equation (F5) of \cite{Gullu4d}. Therefore we shall not directly use these field equations to search for solutions, but instead, we shall employ the symmetry reduction technique in the action before doing calculus of variations. The conventional approach involves inserting the metric into the action and defining the Hubble parameter as $H(t) = \frac{\dot{a}}{a}$ to simplify the Lagrangian into a single equation involving the Hubble parameter and its time derivatives. However, it is crucial to exercise caution in this reduction process, because the "integrated-out" symmetry group in this case is not compact and, therefore, it falls outside the scope of Palais' symmetric criticality guarantee \cite{Palais, short}. Consequently, merely varying the action concerning the scale factor or the Hubble parameter provides necessary but insufficient conditions. In this context, we adhere to Weinberg's robust and error-free computation \cite{Weinberg} for guidance. Inserting \eqref{RW} into \eqref{action} and dropping an overall irrelevant constant, we arrive at the reduced action
\begin{equation} \label{reduced_lagrangian}
    I =  \int dt \, a^3 \, {\cal I}( H, \dot H),
\end{equation}
where the scale factor $a$ and the Hubble parameter $H$ depend on the coordinate time $t$. The remarkable property of our theory is that just like Einstein's gravity the second-time derivative of the Hubble parameter does not appear in the reduced action in contrast to other higher derivative theories.  According to our action \eqref{action}, the function ${\cal I}( H, \dot H)$ can be written as
\begin{eqnarray}
    {\cal I}( H, \dot H)= -1 - 4 \lambda_0 + \sqrt{ X_1^3 X_2},
\end{eqnarray}
where $X_1$ and $X_2$ are given as
\begin{equation}
    \begin{aligned}
    & X_1 = 2 \dot h^2 ((c-2)c+4b-2)-2 (c-2) \left(6 h^2+1\right) \dot h+\left(6 h^2+1\right)^2  \, , \\
    & X_2 = 6\dot h^2 (c (3 c+10)-4b-6)+\left(6 h^2+1\right)^2 +6 (c+2)\left(6 h^2+1\right)\dot h \, , 
    \label{X1X2}
    \end{aligned}
\end{equation}
where we used the dimensionless Hubble parameter $h(t) \equiv \sqrt{\gamma} H(t)$ and the dimensionless time $\Tilde{t}\equiv t/\sqrt{\gamma}$, so that derivatives are with respect to dimensionless time in \eqref{X1X2}. 

It is clear that for the RW metric, the Weyl tensor vanishes identically; hence the parameter $a$ does not contribute to the reduced action.  Note also that for the particular case $c=-1$ and $b = -5/2$, one has a purely quartic theory since $X_1 =X_2$ of which the spherically symmetric black hole solutions were studied in \cite{esin}. Equipped with this, the necessary and sufficient equation to solve is the one coming from 
\begin{equation}
    {\frac{ \delta I}{\delta g_{00}}\,\vline}_{RW}  =0,
\end{equation} 
which yields \cite{Weinberg}
\begin{equation}
    \mathcal{I}-H \frac{\partial \mathcal{I}}{\partial H}+\left(-\dot{H}+3 H^{2}\right) \frac{\partial \mathcal{I}}{\partial \dot{H}}+H \frac{d}{dt}\left(\frac{\partial \mathcal{I}}{\partial \dot{H}}\right)=0.
\end{equation}

By inserting the function $\cal{I}$ in  \eqref{reduced_lagrangian} into the last equation, that is the EoM, which is the $00$-component of the corresponding field equations, can be expressed as follows.
\begin{equation} \label{rho_MG}
    \begin{aligned}
    3 H^2 & =  324 \gamma^3  \left( H^8+9 H^4 \dot{H}^2+6 H^3 \ddot{H} \dot{H}- H^2 \dot{H}^3-3 H \ddot{H} \dot{H}^2+\frac{3}{4} \dot{H}^4\right) \\
    & + 108 \gamma^2\left( H^6+\frac{9}{2} H^2 \dot{H}^2+3 H \ddot{H} \dot{H}- \dot{H}^3\right) + \Lambda_0 \equiv \kappa_0 \rho_{\text{MG}} \, ,
    \end{aligned}
\end{equation}
where we have kept the $00$-component of the Einstein tensor on the left, while we defined the rest coming from the modified gravity as $\rho_{\text{MG}}$ and wrote it on the right-hand side, as is often exercised in modified gravity theories. Moreover, from the variation of the action \eqref{action} with respect to the space component of the metric $\delta g_{ii}$ where $i=1,2,3$, or equivalently from the conversation law $\dot{\rho}_{\mathrm{MG}}+3 H\left(\rho_{\mathrm{MG}}+p_{\mathrm{MG}}\right)=0$, one can find the pressure equation.
\begin{equation} \label{p_MG}
    \begin{aligned}
    -\left( 2 \dot H + 3 H^2 \right)  &= -324 \gamma^3\left[ H^8+\frac{8}{3} H^6 \dot{H}+9 H^4 \dot{H}^2+11 H^2 \dot{H}^3 +\frac{1}{12} \dot{H}^4 \right. \\
    & \left. +2 H \ddot{H} \dot{H} \left(6 H^2 +\dot{H} \right)+2 \ddot{H}^2 \left(H^2 - \dot{H} \right)   + \dot{H} \dddot{H} \left( 2H^2  -\dot{H}\right) \right] \\ 
    & -108 \gamma^2\left[ H^6 +2H\dot{H} \left( H^3+\frac{9}{4} H \dot{H}+ 3 \ddot{H} \right) +\ddot{H}^2 +2 \dot{H}^3 +\dot{H} \dddot{H}\right] \\
    & -\Lambda_0 \equiv \kappa_0 p_{\text{MG}}  \, ,
    \end{aligned}
\end{equation}
where we again used the same convention by defining $p_{\mathrm{MG}}$. Then, with the definition of $\rho_{\mathrm{MG}}$ and $p_{\mathrm{MG}}$, the field equations can be expressed in the usual form for the metric \eqref{RW},
\begin{equation} \label{friedmann}
    \frac{3}{\kappa_0}H^2 = \rho_{\text{MG}} \, , \qquad   
    -\frac{1}{\kappa_0}\left( 2 \dot H + 3 H^2 \right) = p_{\text{MG}}  \, ,
\end{equation}
which are called Friedmann equations. Notice that in the de Sitter limit, where the Hubble parameter is a constant $H(t)=H_0$, the equation of state (EoS) parameter tends to approach minus one,
\begin{equation}
    {\omega_{0_ \text{MG}}}  \equiv  \left.{\frac{p_{\text{MG}}}{\rho_{\text{MG}}} \,}\right|_{dS}  = -1 \,
\end{equation}
irrespective of the presence of a bare cosmological constant. The de Sitter solution can be found using the first Friedmann equation in \eqref{friedmann} that yields
\begin{align} \label{h_0_eqn}
    324 H_0^8 \gamma^4+108 H_0^6 \gamma^3-3 H_0^2 \gamma+\Lambda_0 \gamma = 0,
\end{align}
which can be rewritten in dimensionless form as 
\begin{align}
    4 \lambda^{4} + 4 \lambda^{3} - \lambda + \lambda_{0} =0,
\end{align}
where we defined an effective cosmological constant $\Lambda =  3H_0^2 $ and corresponding dimensionless parameters $ \lambda = \gamma \Lambda$ and $\lambda_0 = \gamma \Lambda_0$ for effective and bare one respectively. Note that the last equation is  \eqref{eff_cosmo}. This polynomial equation has four solutions, two of which are imaginary and one of them is not physical in the sense that it gives $\lambda > 1/4$ which corresponds to $\kappa<0$ according to \eqref{eff_newton} so that we have a unique de Sitter solution for given $\lambda_0$. Here, we provide some remarks concerning the free parameters and the higher-order curvature terms within the Lagrangian. As mentioned previously, setting $a=0$, $b=-5/2$, and $c=-1$, the theory assumes a purely quartic nature, resulting in a Lagrangian with a finite number of curvature terms, specifically up to and including $R^4$ order terms. Alternatively, if we opt to keep these parameters as free variables, the conventional approach to derive field equations involves expanding $\operatorname{det}(\delta^\nu_\mu+4\gamma A_\mu^\nu)$ and its square root in the small $\gamma$ limit. Consequently, the Lagrangian incorporates an infinite number of higher-order terms. However, in our analysis, we found that the parameters $a$, $b$ and $c$ do not appear in the de Sitter solution; and terms beyond the fourth order in curvature exhibit no influence on both the de Sitter solution and the linear perturbations around it. This observation implies that, when considering the simplified cosmological scenarios of a de Sitter universe and the quasi-de Sitter analysis, it is sufficient to focus on the lower-order curvature terms, thereby we can safely set the parameters to the desired values. In that case, the Lagrangian in \eqref{action} can be reduced to the simpler form \footnote[1]{Note that this is the reduced form of the action \eqref{action} with parameters $a$, $b$, $c$ fixed, and this action is not dictated by the trace anomaly that will be introduced in the next section.}.
\begin{equation}
    \mathcal{L}=\frac{1}{2 \gamma \kappa_0}\left( \gamma \left(R-2 \Lambda_0 \right) + \gamma^2 \frac{9}{2} \mathcal{G} + \gamma^3 \frac{1}{2}\left(9\mathcal{G} R-R^3 \right) + \gamma^4 \frac{1}{8} \left( R^4- 18 \mathcal{G} R^2+81\mathcal{G}^2 \right) \right) \,,
    \label{quartic_lagrangian}
\end{equation}
where $\mathcal{G} \equiv R_{\mu \nu \sigma \rho}R^{\mu \nu \sigma \rho}- R_{\mu \nu}R^{\mu \nu} + R^2$ is the Gauss-Bonnet invariant. Then the BI gravity reduces to the form a $\mathcal{F}(R,\mathcal{G})$ theory with the action
\begin{equation}\label{fRG_action}
    I=\frac{1}{2 \kappa_0} \int d^4 x \sqrt{-g} \mathcal{F}(R, \mathcal{G}) \,,
\end{equation}
with a specific $\mathcal{F}$ given as 
\begin{equation} \label{fRG}
    2 \gamma \mathcal{F} \equiv\left(1+\gamma R-\frac{1}{2} \gamma^2\left(R^2-9 \mathcal{G}\right)\right)^2-4 \gamma \Lambda_0-1 \, .
\end{equation}
This is the {\it minimal quartic extension of Einstein's gravity} in the sense that it has the same particle spectrum and a unique vacuum. With our parameter settings, the explicit form of Lagrangian \eqref{quartic_lagrangian} shows that it is quartic in curvature. Henceforth, in the remainder of the analysis, we shall adopt this simpler theory. In principle, one would like to study the process of inflation within the BI gravity described so far, but it turns out the theory does not have a quasi-de sitter phase. Such an attempt was made by one of us in \cite{Tekin_inflation} expecting that the theory would start from the unstable $\kappa <0$, vacuum, that is the repulsive gravity phase, and end up in the $\kappa >0$ vacuum in an inflated state. But, due to a computational error in that work, the claimed result is not correct. Although both Einstein's gravity and our extended version admit maximally symmetric vacuum solutions, they fail to generate an early inflationary phase on their own. It turns out that the contributions from the matter sector,  specifically the trace anomaly arising from the quantum corrections of matter fields in our case, are crucial for realizing an inflationary phase. Without the matter part, our extended theory based solely on the Born-Infeld gravitational action \eqref{action} does not provide a viable quasi-de Sitter inflationary solution, much like the ordinary Einstein's theory without additional fields,  $G_{\mu\nu} = 0$.  Hence, we must resort to the trace anomaly to realize inflation. We describe some basics of the trace anomaly in the next section before applying it to quartic theory. 

%%%%%%%%%%%%%%%%%%%%%%%%%%%%%%%%%%%%%%%%%%%%%%%%%%

\section{Trace anomaly driven inflation in the minimal quartic extension of Einstein’s gravity \label{TA_section}}

\subsection{Trace anomaly}
In one of the earliest models of inflation \cite{Starobinsky}, the expansion is driven by the trace anomaly arising from a significant number of matter fields. The standard model of particle physics encompasses almost a hundred such fields, and this count at least doubles if we consider embedding the standard model within a supersymmetric framework. Consequently, a substantial number of matter fields existed in the early universe. At the energy scale of inflation, these fields can be considered as non-interacting and massless. Within this approach, we conduct a path integral over the matter fields, $\Phi$, within a specified background $g_{\mu\nu}$, yielding an effective action that is functional of the background metric:
\begin{equation}
    e^{i \Gamma\left(g_{\mu \nu}\right)}=\int \mathcal{D} \Phi e^{i S\left(\Phi, g_{\mu \nu}\right)} \,,
\end{equation}
where the classical action can be divided into vacuum and matter parts, $S(\Phi,g_{\mu\nu})=S_\text{vac}(g_{\mu\nu})+S_\text{matter}(\Phi,g_{\mu\nu})$. For a viable theory that is renormalizable, in addition to the Einstein-Hilbert action, the vacuum part should include four different four-derivative terms,
\begin{equation}
    S_{HD}= \int d^4 x \sqrt{g}\left (b_1 W + b_2 \mathcal{G}+b_3 \square R + b_4 R^2\right )\,,  \end{equation}
which are called higher derivative terms that arise naturally in renormalization \cite{Stelle}. Then, the vacuum action is given by
\begin{equation}
    S_\text{vac}=S_{bac.}+S_{HD}\,.
\end{equation}
where $S_{bac}$ is the action of background gravitational model. Furthermore, the effective action of gravity $\Gamma[g_{\mu\nu}]$ admits a loop (or $\hbar$) expansion
\begin{equation}
    \Gamma\left[g_{\mu \nu}\right]=S_{v a c}\left[g_{\mu \nu}\right]+\bar{\Gamma}^{(1)}+\bar{\Gamma}^{(2)}+\bar{\Gamma}^{(3)}+\ldots,
\end{equation}
so that one can write the total expression for the divergent part of the one-loop effective action of the vacuum for the theory involving $N_S$ real scalars, $N_F$ Dirac spinors and $N_V$ massless vectors
\begin{equation}
    \bar{\Gamma}^{(1)}_{d i v}=\frac{1}{\varepsilon} \int d^4 x \sqrt{g}\left (\beta_1 W - \beta_2 \mathcal{G}+\beta_3 \square R + \beta_4 R^2 \right ),
\end{equation}
where $\beta$s are renormalization group $\beta$-functions that depend on the number of fields present in the theory; and given as \cite{shapiro2008,hawking2001} 
\begin{equation}
    \begin{aligned}
    & \beta_1=\frac{1}{120(4 \pi)^2}\left(N_{\mathrm{S}}+6 N_{\mathrm{F}}+12 N_{\mathrm{V}}\right), \,\,\,\,
    \beta_2=\frac{1}{360(4 \pi)^2}\left(N_{\mathrm{S}}+11 N_{\mathrm{F}}+62 N_{\mathrm{V}}\right), \\
    & \beta_3=\frac{1}{180(4 \pi)^2}\left(N_{\mathrm{S}}+6 N_{\mathrm{F}}-18 N_{\mathrm{V}}\right), \,\,\,\,
    \beta_4 = \frac{1}{2 (4 \pi)^2} N_{\mathrm{S}} \left(\xi - \frac{1}{6} \right)^2. \\
    \end{aligned}
\end{equation}
Here, $\xi$ is the non-minimal coupling parameter for the term $\xi R \phi^2$ of a scalar field coupled to gravity, and the value $\xi=1/6$ corresponds to the special version of the scalar theory which possesses local conformal symmetry. Thus, in the conformal case, one can set $\beta_4 = 0$, then the action $S_{HD}$ satisfies the conformal Noether identity 
\begin{equation}
    -\frac{2}{\sqrt{-g}} g_{\mu \nu} \frac{\delta S_{H D}}{\delta g_{\mu \nu}}=0\,,
\end{equation}
which means zero trace for the stress tensor of the vacuum $T^\mu_\mu=0$. At the quantum level, this condition is violated by an anomaly, the trace anomaly that reads as 
\begin{equation}\label{trace_anlomaly}
    \left\langle T_\mu^\mu\right\rangle=-\frac{2}{\sqrt{-g}} g_{\mu \nu} \frac{\delta \bar{\Gamma}^{(1)}}{\delta g_{\mu \nu}} = \alpha' W - \beta \mathcal{G} + \delta \square R \,,
\end{equation}
where $\alpha'=\beta_1$, $\beta=\beta_2$, and $\delta=\beta_3$ for global conformal symmetry. Here, the emergence of the $\Box R$-term in the trace of the energy-momentum tensor is important since this is the term responsible for the Starobinsky instability. We should note that the coefficients $\beta_1$ and $\beta_2$ are independent of the renormalization scheme, but $\beta_3$ is not. For example, the notion adopted here that is the result given by the zeta-function regularization or predicted by AdS/CFT gives $-18$ as the coefficient of $N_V$ in the expression of $\beta_3$ while the result given by dimensional regularization has $+12$. The renormalization-dependent nature of $\delta$ suggests the possibility of adding a finite $R^2$ counter-term to the action at the classical level. When $\mathcal{N}=4$ super Yang-Mills theory is considered, for example, the field content can be given by a single parameter $N$ which is a large number. In that case, the effect of $\Box R$ vanishes and then the inflation never ends. However, $\delta$ can be adjusted to any desired value by adding the finite counter-term.
\begin{equation}
    S_{ct}=\frac{N^2\delta}{12 (4\pi)^2} \int_{\mathcal{M}} d^4 x \sqrt{-g} R^2 \,.
\end{equation}
This counter-term explicitly breaks conformal invariance. In the literature, there is an ambiguity in the parameter $\delta$ and a comprehensive examination of this matter is available in  \cite{shapiro2008,shapiro2006,duff1993}. The presence of $S_{ct}$ implies that we are effectively dealing with a higher derivative theory of gravity, but it is arbitrary to consider it as part of the gravitational action or as part of the matter action. Since we are using the Born-Infeld action as the gravitational sector, the origin of the parameter $\delta$ can be regarded as a matter contribution. Hence, we provide particle content of three example models in Table \ref{tab:particlecontents}, and adopted the Minimum Standard Model in our analysis mainly.

\begin{table} \small	
\captionsetup{format=plain, font=footnotesize, labelfont=bf}
    \centering
    \begin{tabular}{|c|c|c|c|c|}
        \hline
        \textbf{Model} & $N_S$, $N_F$, $N_V$ & $\beta_1$ & $\beta_2$ & $\beta_3$ \\
        \hline 
        \hline
        MSM & (4, 24, 12) & $73/(480\pi^2)$ & $253/(1440\pi^2)$ & $-17/(720\pi^2)$ \\
        \hline
        MSSM & (104, 32, 12) & $11/(48\pi^2)$ & $5/(24\pi^2)$ & $1/(36\pi^2)$ \\
        \hline
        $SU(\mathcal{N})$ SYM & (6$N^2$, 2$N^2$, $N^2$), $N\gg 1$ & $N^2/(64\pi^2)$ & $N^2/(64\pi^2)$ & 0 \\
        \hline
    \end{tabular}
    \caption{Particle contents of Minimum Standard Model (MSM), Minimal Supersymmetric Standard Model (MSSM) and $SU(\mathcal{N})$ Super Yang-Mills Theory, and corresponding values of $\beta$-functions.}
    \label{tab:particlecontents}
\end{table}

Finally, the scenario can be succinctly encapsulated in the following manner: In curved space-time, the action governing the massless matter fields, that are scalars, Dirac spinors, and vectors, exhibits conformal invariance, yet the presence of one-loop vacuum contributions introduces certain divergences. Within the context of renormalization, specific counter-terms are introduced to remove the singularities within the divergent part, albeit at the cost of breaking the conformal invariance inherent to the matter action itself. Classically, within a conformally invariant theory, the trace of the energy-momentum tensor is expected to be zero. However, the renormalization process results in the emergence of an anomalous trace of the energy-momentum tensor in \eqref{trace_anlomaly}, commonly referred to as the conformal anomaly or the quantum anomaly.

%%%%%%%%%%%%%%%%%%%%%%%%%%%%%%%%%%%%%%%%%%%%%%%%%%

\subsection{Trace anomaly within minimal quartic theory}

The final action in the presence of quantum contributions (QC),
\begin{equation}
    I=\frac{1}{2 \kappa_0} \int_{\mathcal{M}} d^4 x \sqrt{-g} \left ( \mathcal{F}(R,\mathcal{G}) +   2\kappa_0 \mathcal{L}_{QC}\right)
\end{equation}
has the field equations:
\begin{equation} \label{FieldEquations}
    \begin{aligned}
    \mathcal{E}_{\mu \nu} \equiv & \mathcal{F}_R R_{\mu \nu}+\frac{1}{2} g_{\mu \nu}\left(\mathcal{G} \mathcal{F}_{\mathcal{G}}-\mathcal{F}\right)+\left(g_{\mu \nu} \square-\nabla_\mu \nabla_\nu\right) \mathcal{F}_R \\
    & +4\left[\left(2 C_{\mu \sigma \nu \lambda}-R_{\mu \sigma \nu \lambda}\right) \nabla^\sigma \nabla^\lambda+\frac{R}{6}\left(g_{\mu \nu} \square-\nabla_\mu \nabla_\nu\right)\right] \mathcal{F}_{\mathcal{G}}= \kappa_0 \left\langle T_{\mu \nu} \right\rangle
    \end{aligned}
\end{equation}
where $\mathcal{F}$ is defined in \eqref{fRG}, and $\mathcal{F}_R$ and $\mathcal{F}_\mathcal{G}$ are partial derivative of $\mathcal{F}(R,\mathcal{G})$ with respect to $R$ and $\mathcal{G}$. The energy-momentum tensor $\left\langle T_{\mu \nu} \right\rangle=\left\langle 0 \left| T_{\mu \nu} \right| 0 \right\rangle $ emerges from the vacuum expectation value of the quantum corrections. For a further discussion on the trace anomaly, see \cite{Gurses_Tekin} and the references therein.

Then, by taking trace of the field equations, we have 
\begin{equation}
    R \mathcal{F}_R+2 \mathcal{G} \mathcal{F}_{\mathcal{G}}-2 \mathcal{F}+3 \square \mathcal{F}_R-4 G_{\mu \nu} \nabla^\mu \nabla^\nu \mathcal{F}_{\mathcal{G}}=\kappa_0 \left\langle T_\mu^\mu\right\rangle \,.
\end{equation}

In the flat RW background with the metric \eqref{RW}, let us build the energy-momentum tensor as
\begin{equation}
    \left\langle T_{00}\right\rangle=\rho, \quad\left\langle T_{i j}\right\rangle=p a(t)^2 \delta_{i j}, \quad(i, j=1,2,3).
\end{equation}
Then, the field equations can be written in a simple form
\begin{equation} \label{EoM}
    \begin{array}{r}
    \frac{3}{\kappa_0} H^2= \rho_{\mathrm{MG}} + \rho  \equiv \rho_{\mathrm{eff}}, \\
    -\frac{1}{\kappa_0}\left(2 \dot{H}+3 H^2\right)=p_{\mathrm{MG}} +  p   \equiv p_{\mathrm{eff}},
    \end{array}
\end{equation}
where $\rho_{\mathrm{MG}}$ and $p_{\mathrm{MG}}$ are previously defined in Eqs. \eqref{rho_MG} and \eqref{p_MG}, and $\rho_{\mathrm{eff}}$ and $p_{\mathrm{eff}}$ are the effective energy density and pressure. In other words, these are the total energy constituents of the universe, comprising quantum matter components and modification of gravity. Furthermore, considering the covariant derivative of the field equations \eqref{FieldEquations}, $\nabla^\mu \mathcal{E}_{\mu 0} = 0$, one arrives at the conservation law.
\begin{equation}
    \label{effective_conservation}
    \dot{\rho}+3 H\left(\rho+p\right)=0 \,.
\end{equation}
Now, we can obtain the full effective energy-momentum tensor. In principle, the complete structure of the energy-momentum tensor originating from the quantum corrections remains unknown. What we do possess knowledge of is its trace, denoted as $\left\langle T_\mu^\mu\right\rangle$. Nonetheless, when provided with the metric, it is possible to derive the energy-momentum tensor of the quantum anomaly on-shell. From the trace $\left\langle T_\mu^\mu\right\rangle$, we have
\begin{equation}
    -\rho+3 p \equiv -\beta \mathcal{G}+\delta \square R, 
\end{equation}
with $W = 0$ in flat RW metric. Using the conservation law in  \eqref{effective_conservation} and eliminating the pressure, we can write
\begin{equation}
    \begin{aligned}
    \frac{d}{d t}\left(\rho a^4\right)=  -\dot{a} a^3\left(-\rho+3 p\right) =
     a^3 \dot{a}\left (24 \beta \frac{\dot{a}^2 \ddot{a}}{a^3}+\delta(\ddot{R}+3 H \dot{R})\right ).
    \end{aligned}
\end{equation}
The integration of this equation provides us the expression for the energy density
\begin{equation} \label{rho_QC}
    \rho=  \frac{\rho_0}{a^4}+6 \beta H^4+\delta\left(18 H^2 \dot{H}+6 \ddot{H} H-3 \dot{H}^2\right), 
\end{equation}
where $\rho_0$ is the integration constant. Using the conservation law, eventually one can find the pressure
\begin{equation}
    p=  \frac{\rho_0}{3 a^4}-\beta\left(6 H^4+8 H^2 \dot{H}\right)-\delta\left(9 \dot{H}^2+12 H \ddot{H}+2 \dddot{H}+18 H^2 \dot{H}\right). 
\end{equation}
Note that the presence of $\rho_0$ implies that the quantum state may contain an arbitrary amount of radiation. However, we can reasonably disregard this radiation due to our focus on extremely high energy scales, all the way to the Planck scale. At this scale, the energy density associated with the standard radiation is typically insignificant compared to that of quantum corrections and gravitational modifications. Consequently, we set $$\rho_0 = 0\,.$$

To study the cosmological evolution of the model, it suffices to use one of the field equations \eqref{EoM}. In what follows, we use the first one, where $\rho_{\mathrm{MG}}$ is defined in \eqref{rho_MG} and $\rho$ is found in  \eqref{rho_QC} with $\rho_0 = 0$.

%%%%%%%%%%%%%%%%%%%%%%%%%%%%%%%%%%%%%%%%%%%%%%%%%%

\subsubsection{de Sitter solutions}
Let us recast the EoM which is the first equation in the Friedmann equations \eqref{EoM} in full form. 
\begin{equation} \label{mastereqn}
    \begin{aligned}
    3 H^{2} = & \Lambda_{0}  + 3 \kappa_{0} \left(2 H^{4} \beta + \delta \left(6 H^{2} \dot H + 2 H \ddot H - \dot H^{2}\right)\right)  \\ &+ \gamma^{3} \left(324 H^{8} + 2916 H^{4} \dot H^{2} + 1944 H^{3} \ddot H \dot H - 324 H^{2} \dot H^{3} - 972 H \ddot H \dot H^{2} + 243 \dot H^{4}\right) \\ &+ \gamma^{2} \left(108 H^{6} + 486 H^{2} \dot H^{2} + 324 H \ddot H \dot H - 108 \dot H^{3}\right).
    \end{aligned}
\end{equation}
Here, the terms with $\kappa_0$ are the effect of trace-anomaly while the effect of modified gravity appears as the terms multiplied by $\gamma$. Solution for a constant $H(t)=H_0$ gives us de Sitter solution which is the solution of the resultant polynomial equation,
\begin{equation}\label{dSeqn}
    \begin{aligned} 
    324 H_{0}^{8} \gamma^{3} + 108 H_{0}^{6} \gamma^{2} + 6 H_{0}^{4} \kappa_{0} \beta - 3 H_{0}^{2} + \Lambda_{0} = 0 \,, \\
    \frac{2 \beta_{0} \lambda^{2}}{3} + 4 \lambda^{4} + 4 \lambda^{3} - \lambda + \lambda_{0} =0 \, ,
    \end{aligned}
\end{equation}
where we defined $\beta_0 = \frac{\beta \kappa_0}{\gamma}$ and $\delta_0 = \frac{\delta \kappa_0}{\gamma^2}$ to be used later. The second line of \eqref{dSeqn} shows that only the Gauss-Bonnet term in the anomalous trace contributes to the de Sitter solution, on the other hand, we will see that the coefficient $\delta$ in front of $\square R$ term plays a crucial role in stability analysis of de Sitter solution. Here, the appearance of a term of $\lambda^2$ order makes a double de Sitter solution possible in the physical range $\kappa>0$, depending on the values of the parameters. One of them is due to the presence of the bare cosmological constant $\Lambda_0$, and the other is due to the trace anomaly, and if the latter is unstable it corresponds to inflation.

%%%%%%%%%%%%%%%%%%%%%%%%%%%%%%%%%%%%%%%%%%%%%%%%%%

\subsubsection{Perturbations around the de Sitter solution}

To find the stability condition of the de Sitter solution, we analyze the linearized perturbations around it $H(t)=H_0 + \Delta H(t)$. Substituting this into \eqref{mastereqn}, and linearizing it in small $\Delta H(t)/ H_0 \ll 1$ gives a solution of the perturbation in the following form,
\begin{equation}\label{deltaH}
    \Delta H(t)=A_0 \, e^{\xi H_0 t},
\end{equation}
where $A_0 \ne 0$ is the amplitude of perturbation; and $\xi$ is the solution of the quadratic equation,
\begin{equation}
    6 H_{0} \left(- 432 H_{0}^{6} \gamma^{3} - 108 H_{0}^{4} \gamma^{2} - 4 H_{0}^{2} \gamma \beta_{0} - H_{0}^{2} \kappa_{0} \delta \xi^{2} - 3 H_{0}^{2} \kappa_{0} \delta \xi + 1\right) =0,    
\end{equation}
which can be expressed in different forms,
\begin{align}
    \xi &= -\frac{3}{2} \pm \frac{ \sqrt{\kappa_{0} \delta \left(- 1728 H_{0}^{6} \gamma^{3} - 432 H_{0}^{4} \gamma^{2} - 16 H_{0}^{2} \gamma \beta_{0} + 9 H_{0}^{2} \kappa_{0} \delta + 4\right)}}{2 H_{0} \kappa_{0} \delta}\nonumber  \\
    &= -\frac{3}{2} \pm \frac{ \sqrt{\delta_{0} \left(- 16 \beta_{0} \lambda + 9 \delta_{0} \lambda - 192 \lambda^{3} - 144 \lambda^{2} + 12\right)}}{2 \delta_{0} \sqrt{\lambda}} \nonumber \\
    &=-\frac{3}{2} \left[ 1 \pm \sqrt{1+\frac{4}{3\delta} \left( \frac{1}{\kappa \Lambda} - \frac{4 \beta}{3} \right)} \, \right].
    \label{xi_soln}
\end{align}
The last expression looks exactly the same as the expression with pure trace-anomaly driven inflation \cite{Bamba2014}, but the difference is that $\Lambda$ is now the effective cosmological constant which is the solution to \eqref{dSeqn} with $ \lambda = \gamma \Lambda$ and $\lambda_0 = \gamma \Lambda_0$, and $\kappa$ is the effective gravitational constant whose relation with the bare $\kappa_0$ and $\Lambda$ is given in \eqref{eff_newton}; so that $\kappa$ implicitly involves $\Lambda$ also. Furthermore, if inflation is described by an unstable de Sitter solution which requires $\xi>0$, the condition for viable inflation can be given as
\begin{equation}
    \kappa \Lambda>0 \, \land \, \beta>\frac{3}{4 \kappa \Lambda} \, \land \, \delta<0.
\end{equation}
Hence, the instability depends on the sign of the $\square R$ term in the trace of energy-momentum tensor. For the particle content of MSM, $\delta$ is negative, while MSSM gives positive values. As discussed above, the renormalization scheme dependence of this term makes it possible to adjust it to any value by introducing a counter $R^2$-term in Lagrangian since having $\Box R$ term in the trace of an energy-momentum tensor and having $R^2$ term in the Lagrangian are equivalent. In Appendix A, we investigate how having this affects the ``potential term'' of the equivalent point-like Lagrangian of our quartic $\mathcal{F}(R,\mathcal{G})$ theory \eqref{quartic_lagrangian}. In most of the analysis, we simply consider the MSM case for the value of $\beta$, and examine negative values of $\delta$ for unstable initial de Sitter phase. 

%%%%%%%%%%%%%%%%%%%%%%%%%%%%%%%%%%%%%%%%%%%%%%%%%%

\subsection{Dynamics of inflation}

\subsubsection{Dynamical system analysis}
\begin{figure}[!ht]%
    \captionsetup{format=plain, font=footnotesize, labelfont=bf}
    \centering
    \subfloat[\centering Einstein's gravity with trace-anomaly. ($\gamma = 0$) ]
    {{\includegraphics[width=7cm]{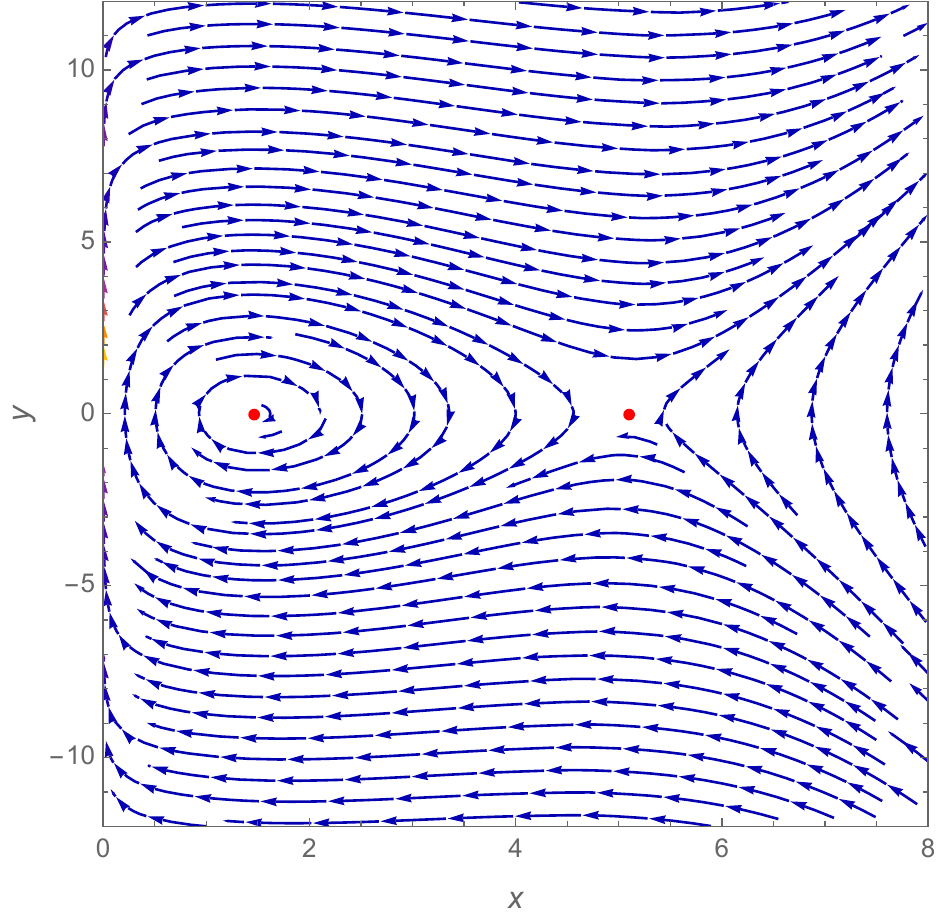} }}%
    \,
    \subfloat[\centering Minimal quartic model without trace-anomaly. ($\gamma = 0.0025$)]
    {{\includegraphics[width=7cm]{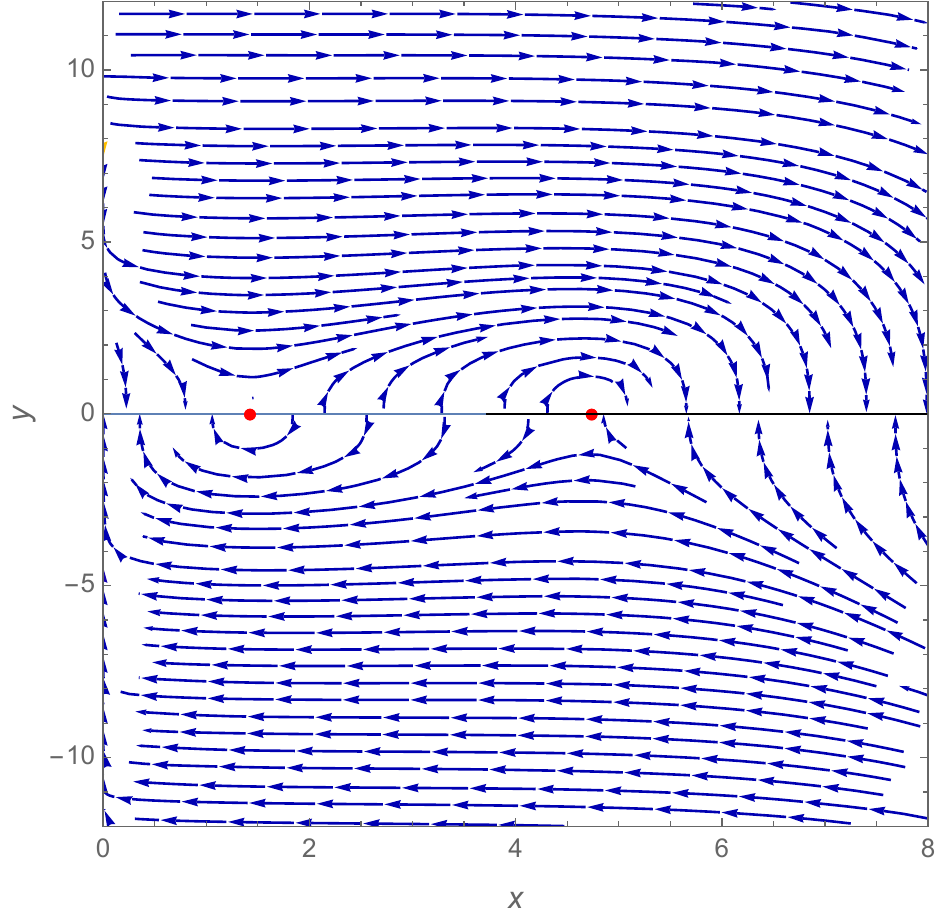} }}%
    \\
    \subfloat[\centering Minimal quartic model with trace-anomaly. ($\gamma = 0.0025$)]
    {{\includegraphics[width=7cm]{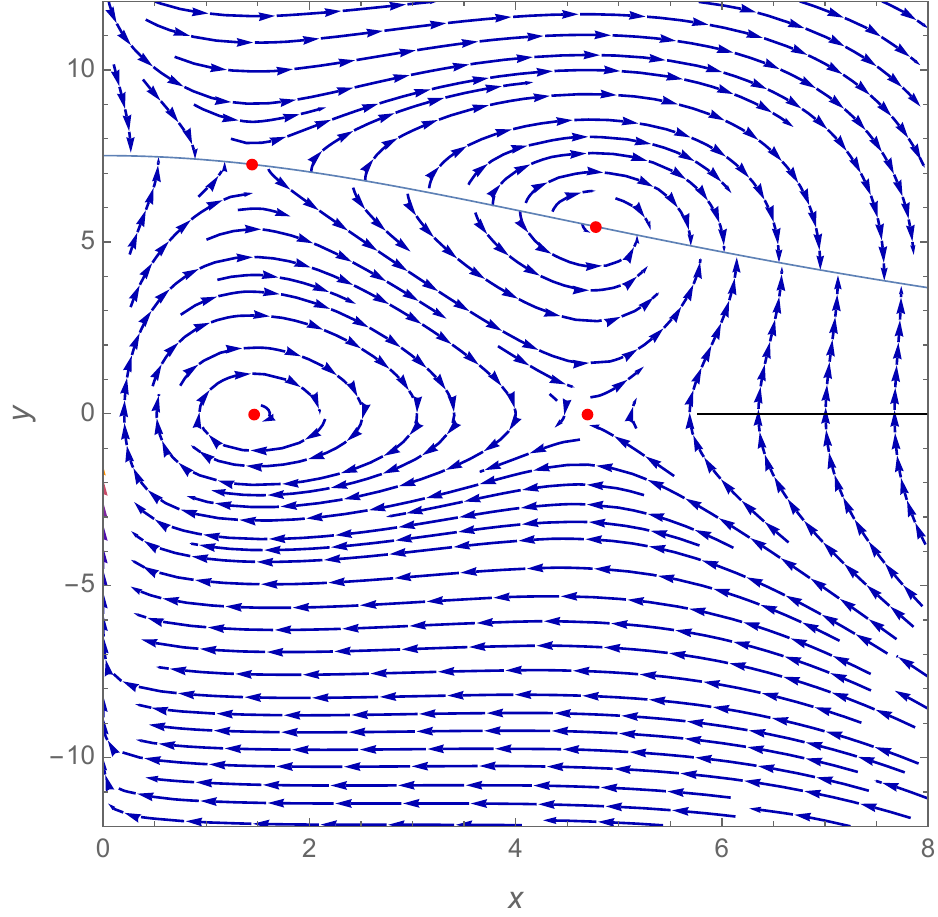} }}%
    \,
    \subfloat[\centering Minimal quartic model with trace-anomaly. ($\gamma = 0.0060$) ]
    {{\includegraphics[width=7cm]{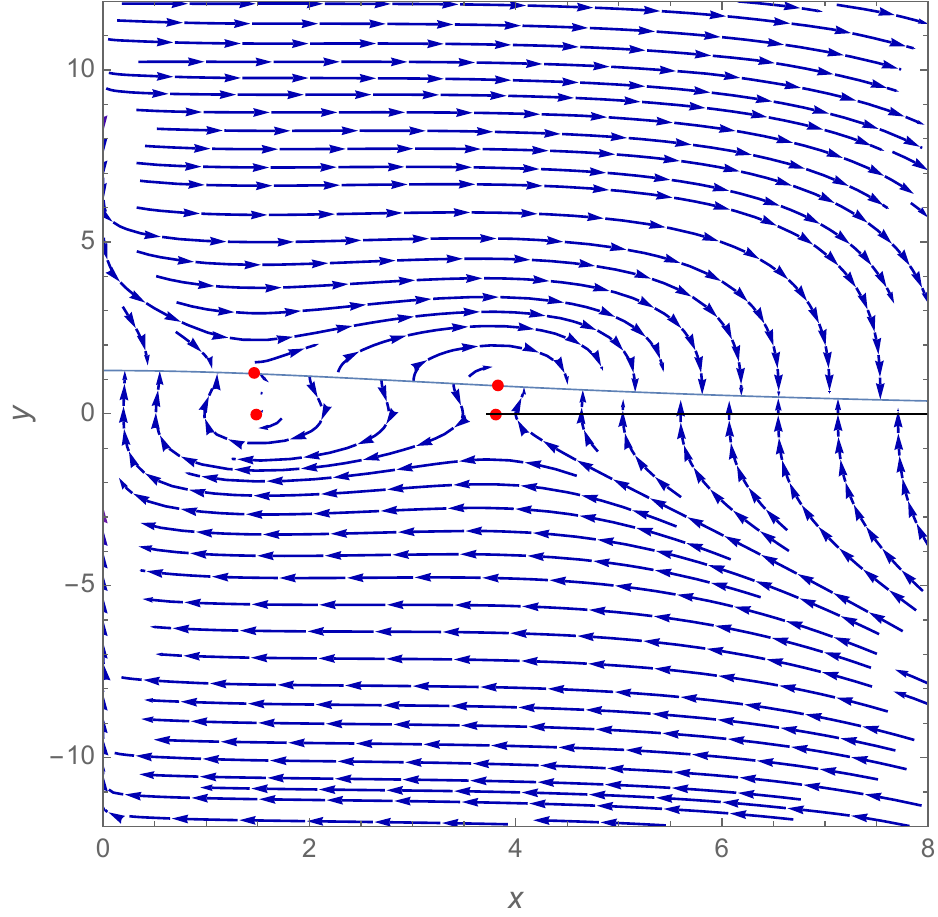} }}%
    \caption{The phase-portrait of the system described by the  EoM \eqref{mastereqn} for the MSM case with parameter settings $x=H(t)$ and $y=\dot H (t)$. The figure is plotted in $c = \hbar = \kappa_0 = 1$ units and the value of the bare cosmological constant is chosen arbitrarily as $\Lambda_0 = 6$. }%
    \label{fig:phaseportrait}%
\end{figure}

We first analyze our main differential equation \eqref{mastereqn} with a phase portrait analysis. This provides us general behavior of the solutions in the phase space without solving the equation. Using a  change of variables, 
\begin{equation}
    x = H(t) \quad \text{and} \quad y=\dot{H}(t), 
\end{equation}
it can be  written as a first order differential equation
\begin{equation}
   \frac{dy}{dx}=f(x,y \, ; \, \Lambda_0,\gamma,\beta,\delta)=\frac{f_1(x,y)}{f_2(x,y)}\,. 
\end{equation}
Here, $\Lambda_0$ and $\gamma$ are free parameters that need to be determined while the other parameters $\beta$ and $\delta$ are determined by the number of particles in the underlying particle theory. The functions $f_1(x,y)$ and $f_2(x,y)$ are
\begin{equation}
    \begin{aligned}
    f_1(x,y) &= \Lambda_0+324 \gamma ^3 x^8+108 \gamma ^2 x^6+6 x^4 \left(\beta  \kappa_0+486 \gamma ^3 y^2\right) \\ &+ x^2 \left(-324 \gamma ^3 y^3+486 \gamma ^2 y^2+18 \delta  \kappa_0 y-3\right)+243 \gamma ^3 y^4-108 \gamma ^2 y^3-3 \delta  \kappa_0 y^2 \\
    f_2(x,y) &= 6 x y \left(\delta  \kappa_0+54 \gamma ^2 y \left(6 \gamma  x^2+1\right)-162 \gamma ^3 y^2\right)
    \end{aligned}
\end{equation}
With our definitions of $x$ and $y$, the velocity field reads 
$$(v_x,v_y)=\left(\frac{dx}{dt}, \frac{dy}{dt} \right)=(\dot H, \ddot H)=(y,y\,f(x,y))\,.$$
Then, plotting this velocity field, one can find and analyze the critical points and their characteristics of the system in $xy$-plane. In Fig.(\ref{fig:phaseportrait}), we provided the phase portrait for the Minimum Standard Model (MSM) in four different regimes.

In the first scenario depicted in Fig.(\ref{fig:phaseportrait}-a), we examine the case where there is no contribution from the modified gravity, representing a pure trace-anomaly situation within the framework of Einstein's gravity. Two critical points can be seen along the $y=0$ axis, corresponding to two distinct de Sitter solutions. For the particle content of MSM, the second solution is identified as unstable. Consequently, the universe is presumed to have originated nearly from this point, having a negative $\dot H$, subsequently transitioning towards the first critical point, which acts as the final attractor of the system, determined by $\Lambda_0$, and may correspond to the late-time acceleration. In the absence of an intrinsic cosmological constant, the ultimate attractor point resides at $(x,y)=(0,0)$. Thus, during the inflationary epoch, the Hubble parameter $H$ experiences a slow decrease while maintaining $\dot H$ in near constancy, eventually oscillating around the final attractor, indicative of the reheating era.

In the second scenario, depicted in Fig.(\ref{fig:phaseportrait}-b), we switch off the trace anomaly and investigate the dynamics of pure modified gravity. Once again, two distinct de Sitter solutions are evident along the $y=0$ axis. Additionally, a solid black line is introduced along the x-axis to indicate a non-physical region where $\lambda>1/4$, a characteristic feature typical of Born-Infeld-type theories, setting an upper limit on the effective cosmological constant. Consequently, the second solution is nonphysical, as it intersects with the solid line, resulting in the existence of a unique de Sitter solution. In this context, a critical surface manifests along the $y=0$ axis, changing the characteristics of critical points both below and above this surface. The unique de Sitter solution remains stable within the domain where $\dot H < 0$, yet transitions to instability when $\dot H > 0$ is encountered. Furthermore, the critical surface is identified as unstable within these points, while exhibiting attractor behavior beyond this region. 

In Fig.(\ref{fig:phaseportrait}-c), we explore our modified gravity theory gently introducing a small value for $\gamma$ in the presence of the trace anomaly. In comparison to the first scenario, the introduction of modified gravity results in the emergence of three critical points situated above the de Sitter solutions observed in the original trace-anomaly case. However, one of these critical points exhibits a notably high $\dot H$ value, surpassing the upper limit of this plot and consequently playing no role in the inflationary solutions. The remaining two critical points reside above the critical points along the $y=0$ axis and are positioned on a critical surface represented by a curve that converges toward the x-axis as $H\rightarrow \infty$. Notably, the critical surface with critical points on it in pure modified gravity case is raised within the $\dot H > 0$ region, while still maintaining the de Sitter solution due to the trace anomaly within the physically plausible range, $\kappa>0$. Under an initial condition where $\dot H \lesssim 0$, the dynamics of inflation remain largely unaffected by the modified gravity.

In Fig.(\ref{fig:phaseportrait}-d), to understand the effect of modified gravity on trace-anomaly-driven inflation, we play with the value of $\gamma$. Increasing $\gamma$ makes the top critical points get closer to the de Sitter points. The solid line indicates the $\kappa<0$ region and, as $\gamma$ increases, it includes the initial de Sitter point which means inflation cannot happen in this (unphysical) case. This imposes an upper bound on $\gamma$ so that the effects of modified gravity should not exceed the trace anomaly for a viable inflationary model. 

\begin{figure}[H]%
    \captionsetup{format=plain, font=footnotesize, labelfont=bf}
    \centering
    \subfloat[\centering Trace-anomaly in Einstein's gravity with MSSM particle content]
    {{\includegraphics[width=7cm]{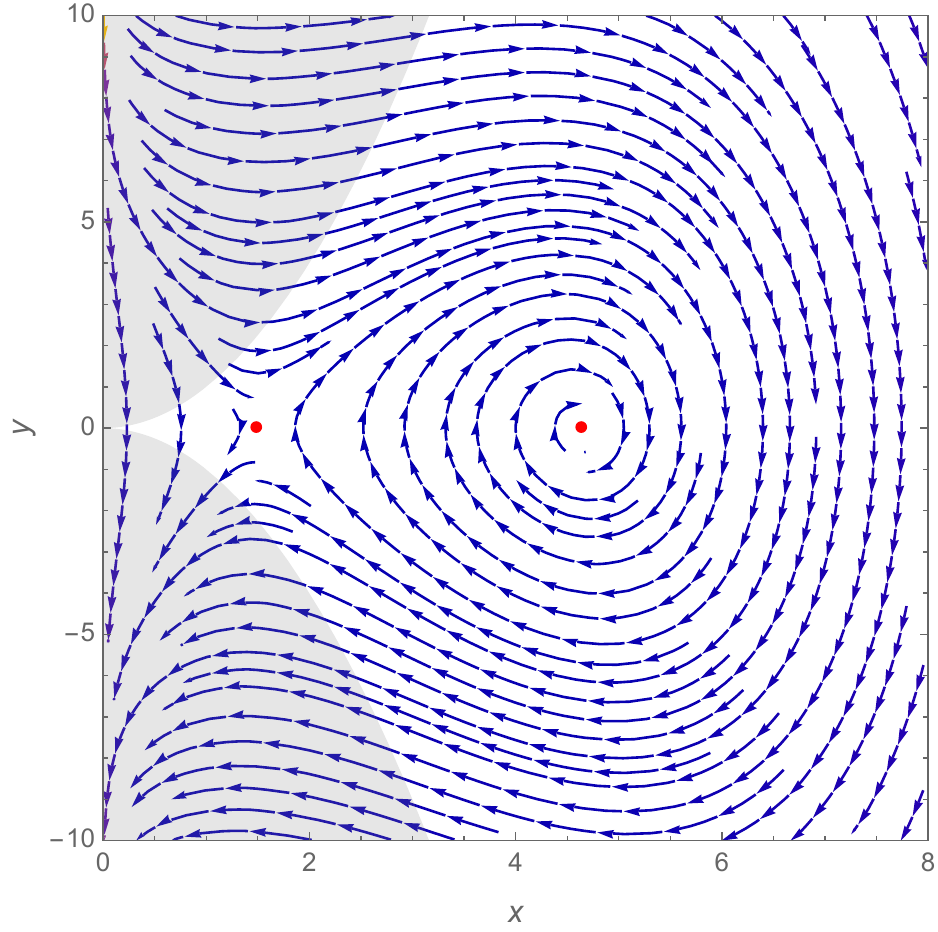} }}%
    \,
    \subfloat[\centering Trace-anomaly in minimal quartic model with MSSM particle content ]
    {{\includegraphics[width=7cm]{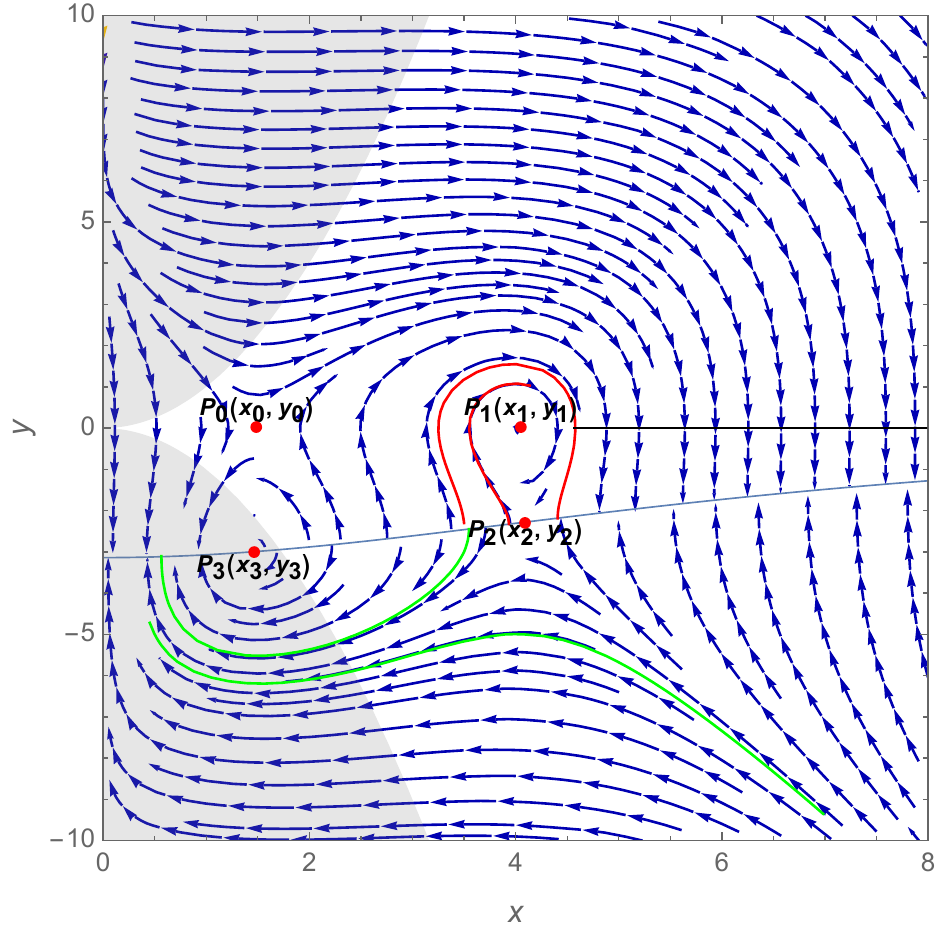} }}%
    \caption{The phase-portrait of the system described by the EoM \eqref{mastereqn} for MSSM case. We shade off the region $|y/x^2|>1$ indicating the outside of the slow-roll regime which is not preferred for initial conditions. The curves highlighted in color in \textbf{(b)}, are some example solutions with (green) and without (red) graceful exit. The figure is plotted in $c = \hbar = \kappa_0 = 1$ units and the parameters are set as $\gamma = 0.004$ and $\Lambda_0 = 6$.}%
    \label{fig:phaseportraitMSSM}%
\end{figure}

Furthermore, we investigate the scenario wherein the sign of the $\Box R$-term is positive, indicative of an initial stable point. This condition finds representation in the context of MSSM as an illustrative example model encompassing the particle content pertinent to this case. In Fig.(\ref{fig:phaseportraitMSSM}), the phase-portrait of the system is plotted without and with the effects of modified gravity for the MSSM case respectively, and from Fig.(\ref{fig:phaseportraitMSSM}-a), it can be seen that the second de Sitter solution due to the trace-anomaly is a stable critical point. In this case, a graceful exit from Inflation requires its mechanism to disrupt this state. In Ref. \cite{starobinsky2016}, a possible transition from stable to unstable phases of inflation is discussed, and different quantum mechanisms that may produce large $R^2$-term during the inflation are proposed for a graceful exit, for example. On the contrary, we show that the minimal extension of Einstein's gravity makes possible an exit from inflation without any other mechanism. In Fig.(\ref{fig:phaseportraitMSSM}-b), we have plotted the contribution of modified gravity in this case. As in MSM case, modified gravity creates additional critical points on the phase-portrait, visible ones of those in the plotted region are labeled as $P_i(x_i,y_i)$ in the graph, but this time the two of them lying on a critical surface are in $\dot H < 0$ region. Note that increasing the intensity of modified gravity makes this critical surface closer to the $y=0$ line. Therefore, if the universe starts with $H \simeq x_1 \simeq x_2$ and a small value of $\dot H$ from below this surface, $\dot H \lesssim y_2$ for example, then the exit from inflation can occur due to the attractor behavior of the surface between $0$ and $x_3$. This prevents $\dot H$ from flowing through negative infinity and ends inflation with $H \lesssim x_3$. If the strength of modified gravity is high enough, the initial point for $\dot H$ can be in the slow-roll regime, i.e. $ \left| \dot H / H^2 \right| \ll 1$ while keeping $\kappa$ positive. In Fig.(\ref{fig:phaseportraitMSSM}-b), the green curves are some examples of possible inflationary solutions with graceful exit while the red ones does not have enough initial $\dot H$ value so that the graceful exit is not possible for those similar to the Einstein's gravity. We should note that the necessary number of e-folding is not guaranteed for all of these solutions, but a particular realistic solution with enough number of e-folding is analyzed in the next section.

%%%%%%%%%%%%%%%%%%%%%%%%%%%%%%%%%%%%%%%%%%%%%%%%%%

\subsubsection{Numerical solutions}
\begin{figure}[H]%
    \captionsetup{format=plain, font=footnotesize, labelfont=bf}
    \centering
    {{\includegraphics[width=11cm]{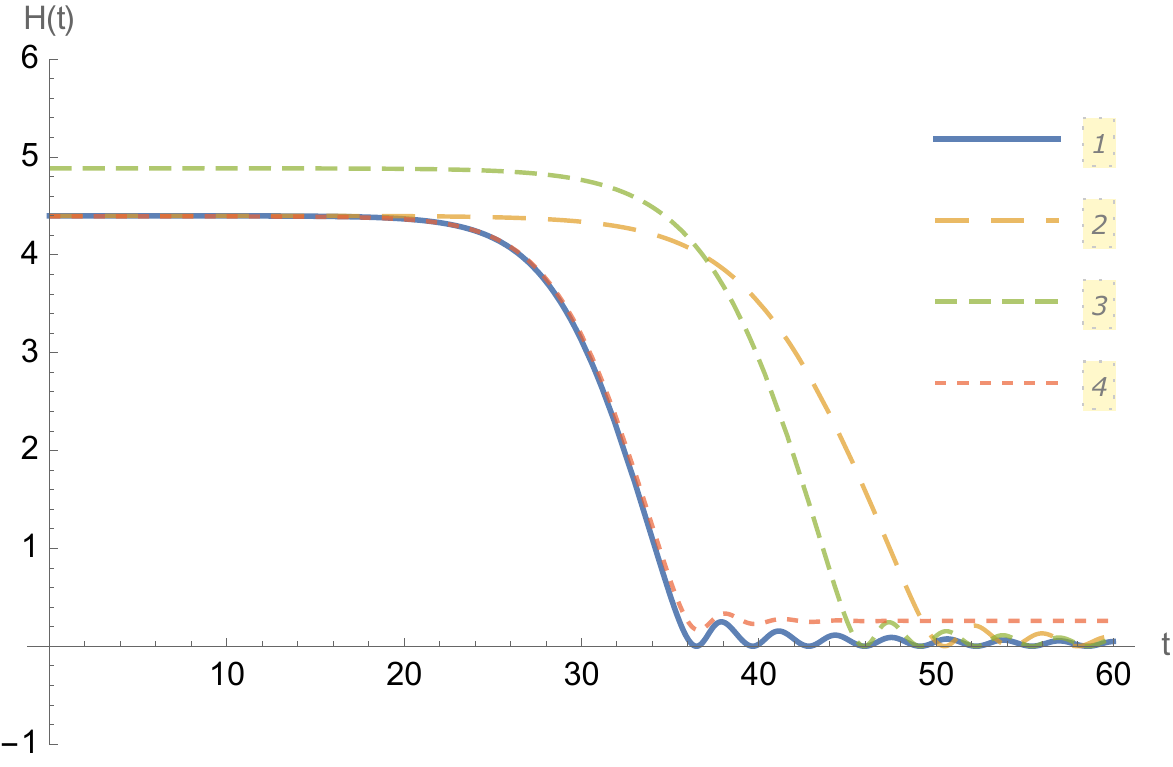} }}%
    \caption{Numerical solutions of $00$-component of field equations for $H(t)$ with different parameter settings in $c = \hbar = \kappa_0 = 1$ units. Here, the value of $\beta=253/(1440\pi^2$ is given by MSM particle content. The other parameters that we used for numerical solutions are given as \\
    {\boxed{\bf{1}}} $\gamma=0.0043$, $\delta=-0.25$, $\Lambda_0=0$;
   {\boxed{\bf{2}}} $\gamma=0.0043$, $\delta=-0.35$, $\Lambda_0=0$;
   {\boxed{\bf{3}}} $\gamma=0.0025$, $\delta=-0.25$, $\Lambda_0=0$;
    {\boxed{\bf{4}}} $\gamma=0.0043$, $\delta=-0.25$; $\Lambda_0=0.2$.
    The initial conditions for the numerical solutions are taken as $H(0)=H_i-10^{-5}$ and $\dot H(0)=0$ where $H_i$ is the solution of \eqref{dSeqn} which is the value of initial de Sitter phase.}
    \label{fig:H_vs_t_numerical}
\end{figure}

Here, we numerically solve EoM which is the $00-$component of the field equations given explicitly in \eqref{mastereqn} for different numerical values of $\gamma$ and $\delta$. Recall that the value of the parameter $\beta$ is fixed by renormalization, but we have a freedom to set $\delta$ to any desired value by adding $R^2$ counter term to the action due to its renormalization scheme dependence. Moreover, because the parameter $\Lambda_0$ is related to the final attractor of the system, we set it to zero $\Lambda_0=0$ for the rest of the analysis, but it can be set to a value for unified early and late time acceleration model. Hence, our main free parameters are BI parameter $\gamma$ and the coefficient of $\Box R$ term, $\delta$, to fit the solution into desired shape. In Fig.(\ref{fig:H_vs_t_numerical}), we give four numerical solution of the field equation for $H(t)$ to judge the model parameters $\gamma$, $\Lambda_0$, and $\delta$, with the parameter $\beta$ set to $\beta=253/(1440\pi^2)$ for the content of the MSM particles.

\begin{figure}[H]%
    \captionsetup{format=plain, font=footnotesize, labelfont=bf}
    \centering
    {{\includegraphics[width=14cm]{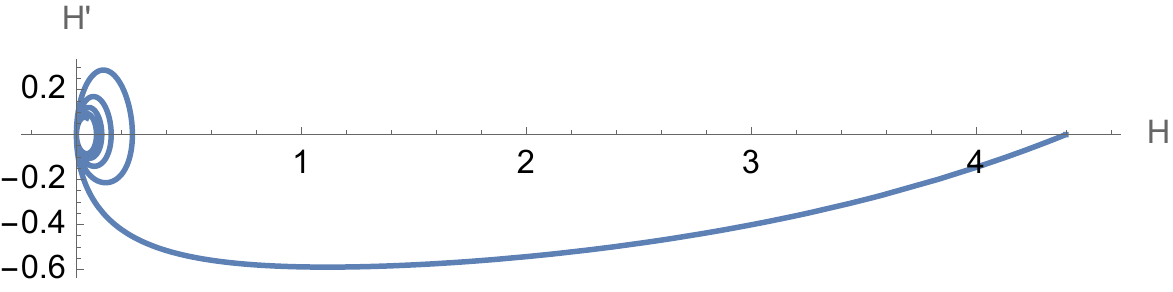} }}%
    \caption{The plot of the numerical solution \boxed{\bf{1}} in $H-\dot H$ phase space. The solution is consistent with the case of phase portrait (\ref{fig:phaseportrait})(c) that we discussed in the dynamical system analysis.}%
    \label{fig:phasespace_num}%
\end{figure}

\begin{figure}[H]%
    \captionsetup{format=plain, font=footnotesize, labelfont=bf}
    \centering
    \subfloat[\centering during inflation]
    {{\includegraphics[width=7cm]{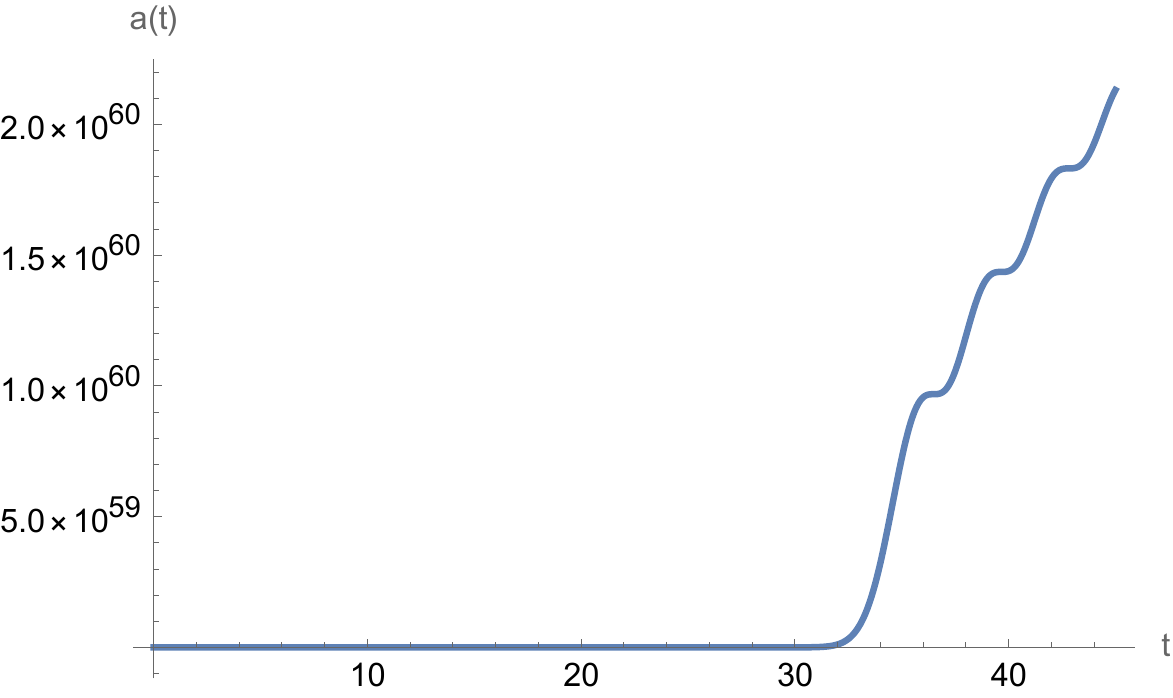} }}%
    \,
    \subfloat[\centering during reheating ]
    {{\includegraphics[width=7cm]{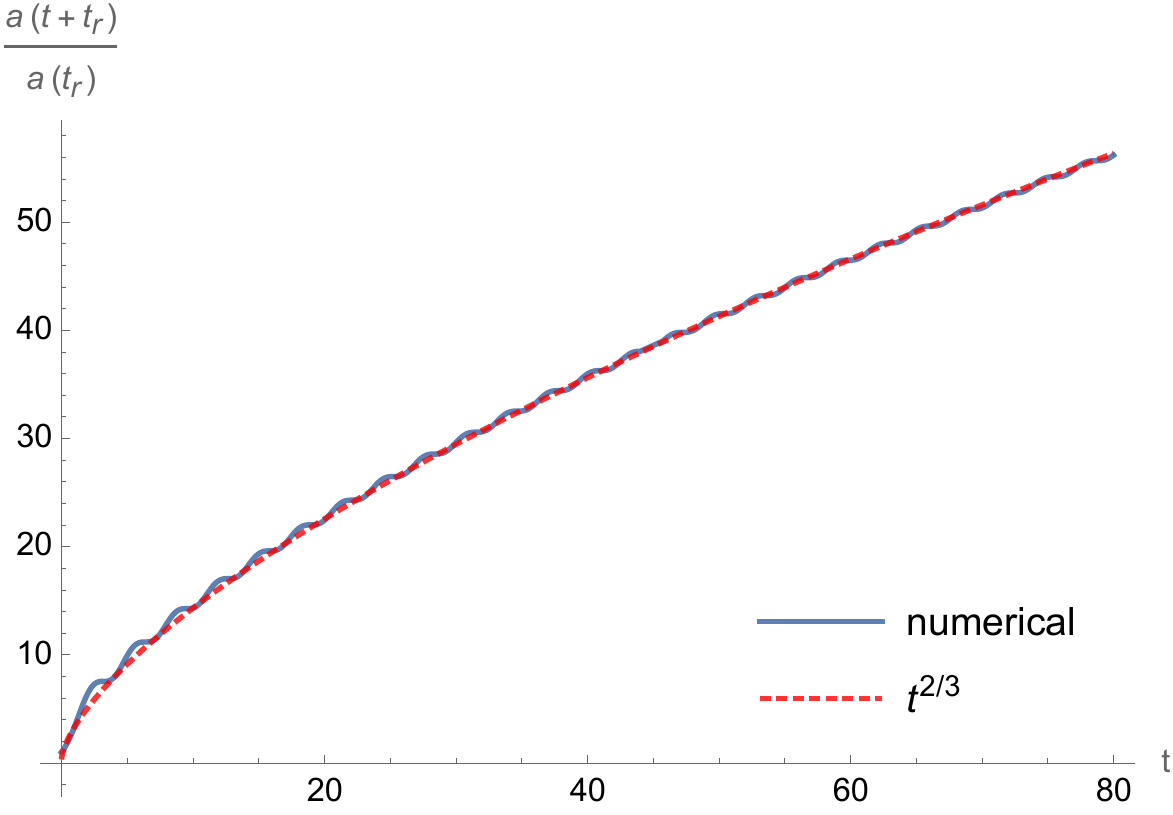} }}%
    \caption{Evolution of the scale factor $a(t)$ in time. In left panel \textbf{(a)}, we see a quasi-exponential expansion which expands space about $10^{59}$ order. On the right panel \textbf{(b)}, the behavior of $a(t)$ during reheating era is presented and shows that the scale factor evolves like matter dominated phase as $a \sim t^{2/3}$. Here, $t_r$ is the initial time of reheating and we re-scaled the initial value by normalizing with $a(tr)$ for this plot.}%
    \label{fig:a_vs_t}%
\end{figure}

\noindent With the help of the numerical results, we have a chance to examine the inflation scenario that we revealed in the phase portrait analysis in more detail. Then, the typical solution of $H(t)$ can be described as follows. $H$ starts with an initial value of the unstable vacuum state determined by $\beta$ and $\gamma$ as a solution of \eqref{dSeqn}, and slowly roll down from this value until it enters a linearly decreasing phase, $H \propto \ -t$, through to zero which is a typical behavior of $R^2$ gravity in the slow roll regime \cite{Mijic1986}. Then, bouncing back from zero, $H$ approaches the value of the final vacuum determined by $\Lambda_0$ by oscillations with decreasing amplitude called the reheating era. If $\Lambda_0=0$, the final value of $H\rightarrow 0$. In Fig.(\ref{fig:phasespace_num}), one of the solutions can be seen in the phase space that we defined in the previous section. In addition, using the numerical solution of $H(t)$, we have also plotted the function $a(t)$ in Fig.(\ref{fig:a_vs_t}), from which the behavior of the scale factor during inflation can be seen as quasi-exponential while during the reheating phase it can be approximated by $a(t)\propto t^\frac{2}{3}$.

\begin{figure}[H]%
    \captionsetup{format=plain, font=footnotesize, labelfont=bf}
    \centering
    {{\includegraphics[width=14cm]{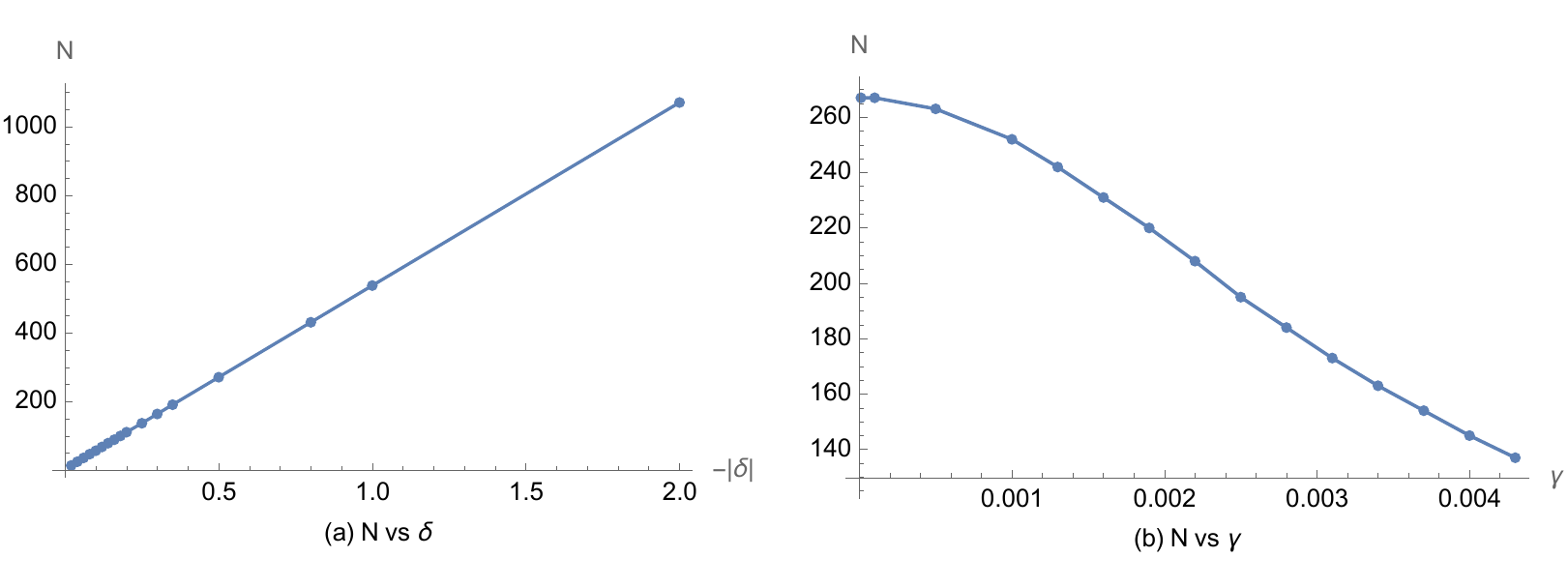} }}%
    \caption{The dependence of total e-folding number $\mathcal{N}$ at the end of inflation on the parameters $\delta$ and $\gamma$. The data points are found by varying model parameters and performing numerous numerical solutions. In \textbf{(a)}, the relation with respect to $\delta$ is calculated at fixed $\gamma=0.0043$, while in \textbf{(b)}, the relation is at fixed $\delta=-0.25$.}%
    \label{fig:multiple_results}%
\end{figure}

In our numerical analysis, the aim is to examined the role of parameters $\delta$ and $\gamma$ on the dynamics of inflation and the number of $e$-folding $\mathcal{N}$.  At the end of inflation $e$-folding number $\mathcal{N}$ is given by
\begin{equation}
    \mathcal{N}\equiv\log{\left[\frac{a(t_f)}{a(t_i)}\right]} \,,
\end{equation}
where $t_i=0$ and $a(t_i)=1$ at the initial time, and $t_f$ is the end of inflation where the Universe enters the oscillatory phase. By varying $\delta$ and $\gamma$ in many numerical solutions, in Fig.(\ref{fig:multiple_results}), we find the relations between $\mathcal{N}$ at the end of inflation and model parameters. Thus, the following observations can be made. First, the duration of the inflation is determined mainly by the term $\delta$, but $\gamma$ also contributes little. A higher $\gamma$ reduces the initial value of the $H$ since it enters the solution of de Sitter equation \eqref{dSeqn} and also it suppress the duration of inflation. However, for fixed $\beta$, there is a maximum value of $\gamma$ at which $\lambda = 1/4$. For MSM particle content, $\beta=253/(1440\pi^2)$, the highest value is $\gamma \simeq 0.00431$ so that a $\gamma$ with higher than that value makes $\kappa$ negative, which is undesired. 

%%%%%%%%%%%%%%%%%%%%%%%%%%%%%%%%%%%%%%%%%%%%%%%%%%

\subsubsection*{The slow-roll parameters}

We can also check if slow-roll parameters during the inflation are small enough. From the second time derivative of $a(t)$,
\begin{equation}
    \frac{\ddot{a}}{a}=H^2+\dot{H}=H^2(1-\epsilon),
\end{equation}
we can introduce the slow-roll parameter $\epsilon$, which has to be small during the inflation to vary $H(t)$ slowly.
\begin{equation}
    \epsilon=-\frac{\dot{H}}{H^2} \ll 1.
\end{equation}
To have a viable inflation with a graceful exit, the slow-roll requires that the change in $\epsilon$ should also be small, so that we have another parameter $\eta$ to measure it.
\begin{equation}
    |\eta|=\left|-\frac{\ddot{H}}{2 H \dot{H}}\right| \equiv\left|\epsilon-\frac{1}{2 \epsilon H} \dot{\epsilon}\right| \ll 1 .
\end{equation}
These two parameters, referred to as the Hubble slow-roll parameters, play a pivotal role in analyzing the feasibility of inflation within the framework of standard inflationary models. In Einstein's gravity with a single scalar field, we can express some of the observables such as spectral index of scalar curvature perturbations and tensor-to-scalar ratio in terms of these parameters. In the modified gravity case, however, there could be more than two slow-roll parameters depending on the structure of the gravity model. In that case, one can introduce parameters as needed, such as \cite{Hwang1996},
\begin{equation}
    \epsilon_1=-\frac{\dot{H}}{H^2}, \quad 
    \epsilon_2=\frac{\ddot{\mathcal{G}}}{H \dot{\mathcal{G}}}, \quad 
    \epsilon_3=\frac{\dot{f}_R}{2 H f_R}, \quad 
    \epsilon_4=\frac{\dot{E}}{2 H E} \quad
    \text{with} \quad E=\frac{3 \dot{f}_R^2}{2 \dot{\mathcal{G}}^2} \,
\end{equation}
and extend this set further if necessary. Then, the true relations between the slow-roll parameters and the observables come from the cosmological perturbations in that theory with scalar vector tensor (SVT) decomposition.

Here, we should make an important remark. Our theory, just like the model of Starobinsky, has a massless spin-2 and a massive spin-0 field, the latter coming from the quantum anomaly. (For the particle content of the model, see Appendix B and also \cite{Tekin3}.) The mass of the scalaron in the Starobinsky case ($\mathcal{L}=R+ \kappa_0 \alpha R^2$), is $m^2_{\phi}=\frac{1}{6 \kappa_0 \alpha}$, while our quartic extension contributes to it via effective gravitational constant so that in the maximally symmetric background, it becomes $m^2_{\phi}=\frac{1}{6 \kappa \alpha}$ where $\kappa$ is given by \eqref{eff_newton}.

\begin{figure}[H]%
    \captionsetup{format=plain, font=footnotesize, labelfont=bf}
    \centering
    \subfloat[\centering Number of $e$-folding ]
    {{\includegraphics[width=7cm]{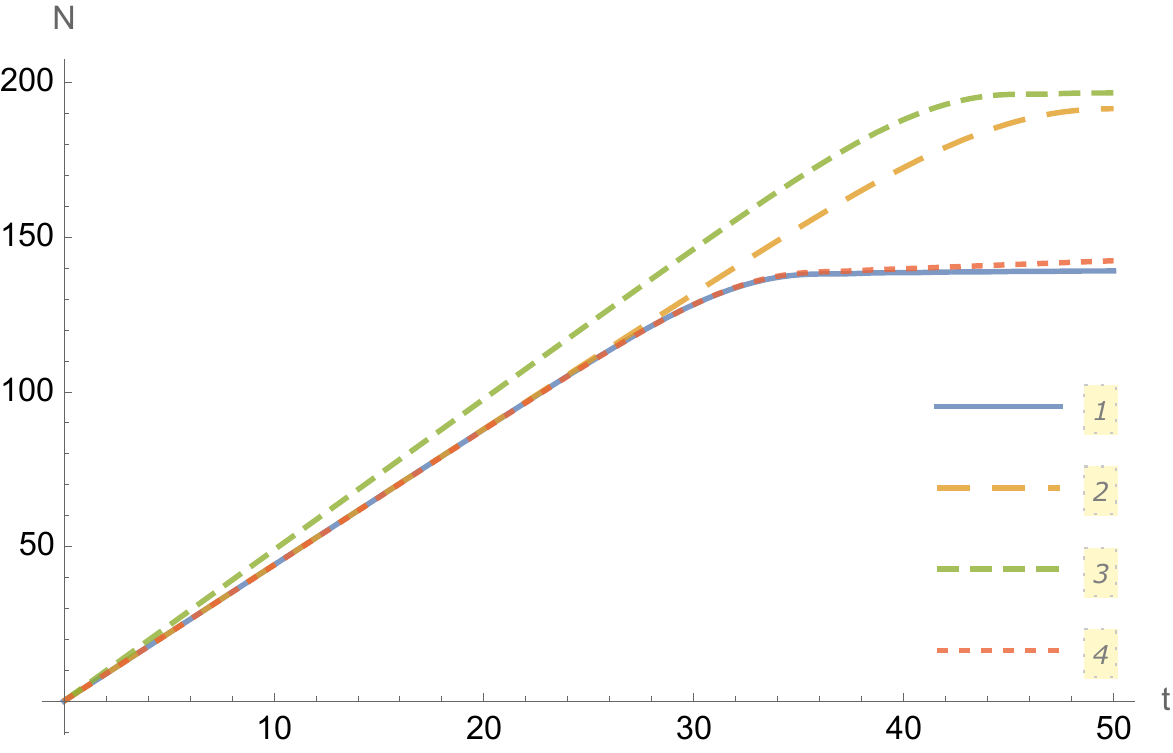} }}%
    \,
    \subfloat[\centering slow roll parameters during inflation]
    {{\includegraphics[width=7cm]{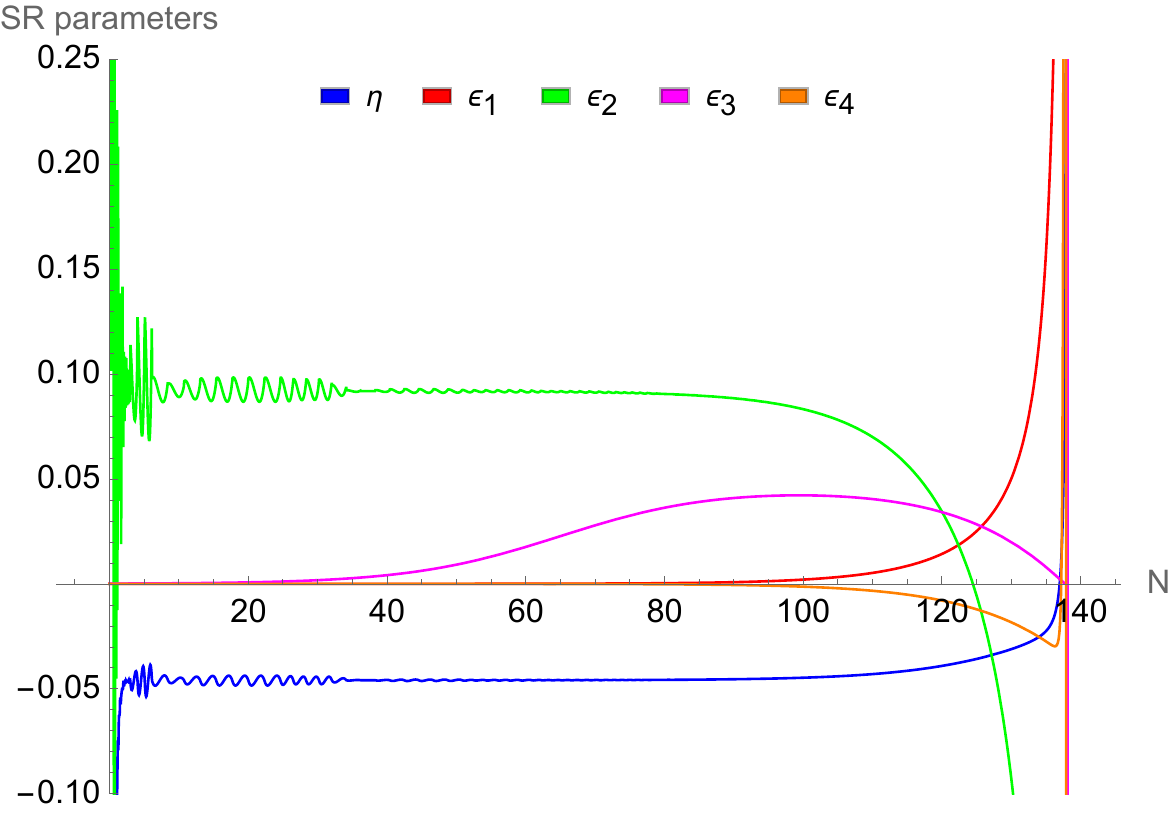} }}%
    \caption{The number of $e$-folding during the inflation with respect to cosmic time t  is given in \textbf{(a)} for different solutions presented in Fig. (\ref{fig:H_vs_t_numerical}). In \textbf{(b)}, The evolution of slow roll parameters $\epsilon_i$ and $\eta$ during the inflation is given with respect to number of $e$-folding $\mathcal{N}$ for the solution 
    {\boxed{\bf{1}}}.%
    }
    \label{fig:ns_numerical}
\end{figure}

In Fig. (\ref{fig:ns_numerical}-b), we have seen that all of these parameters are small during the inflation so that inflationary solutions within this theory fulfill the slow-roll conditions. If the horizon crossing happens at $\mathcal{N}_{*}=50-60$, then the order of these parameters as power of $10$ are given as $\mathcal{O}\left(\eta, \epsilon_1, \epsilon_2, \epsilon_3, \epsilon_4\right)= \{-1,-3,-1,-1,-3\}$. Hence, in the linear approximation with suitable chosen model parameters, any linear combination of these parameters in the expression of spectral index with the coefficients $c_i$, coming from perturbation study, such as $n_s=1-\sum_{i=1}^n c_i \epsilon_i$ can satisfy observational constraints \cite{Planck2018} if $c_i$'s are smaller or around order of 10.

%%%%%%%%%%%%%%%%%%%%%%%%%%%%%%%%%%%%%%%%%%%%%%%%%%

\subsubsection*{Approximate solution}

\begin{figure}[H]%
    \captionsetup{format=plain, font=footnotesize, labelfont=bf}
    \centering
    \subfloat[\centering $H$ vs $t$]
    {{\includegraphics[width=8cm]{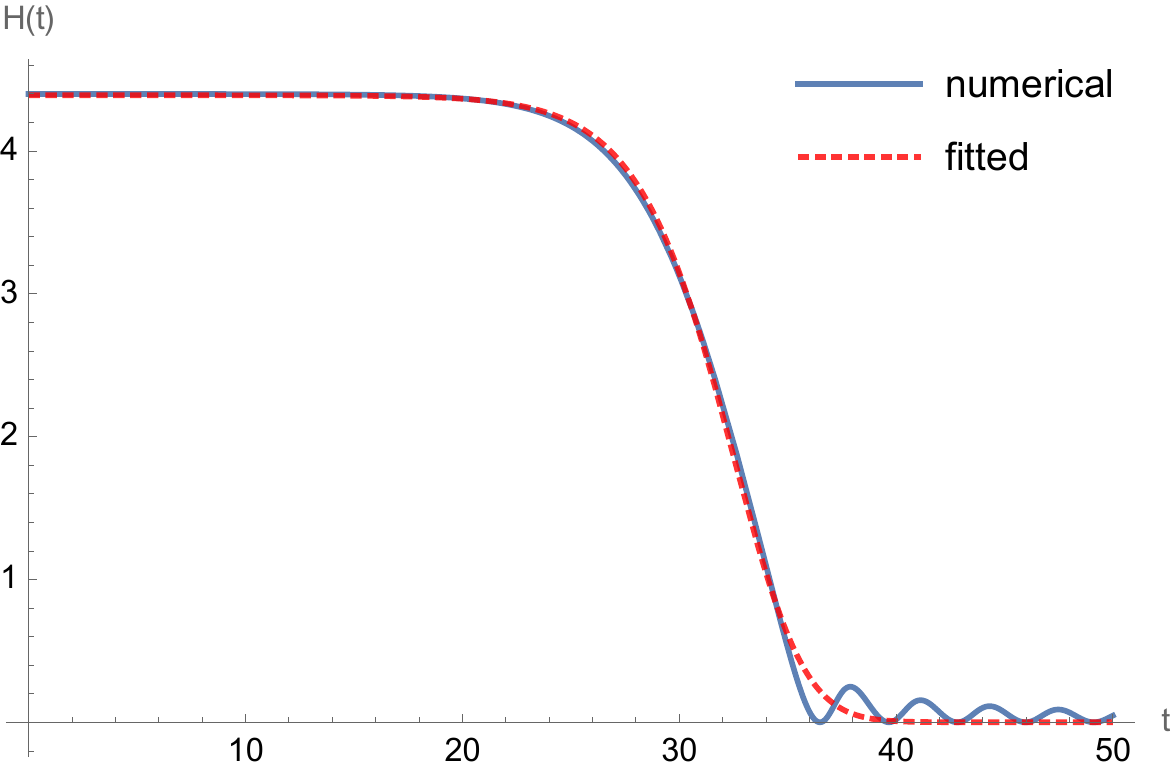} }}%
    \caption{The approximate form of $\bar H(t)$ fitted to the numerical solution {\boxed{\bf{1}}} in Fig.(\ref{fig:H_vs_t_numerical}) with model parameters $\left[H_m, \, k,\, t_0 \right] = \left[4.389,\, 0.414,\, 35.18\right]$ with the value of $n=3$.}%
    \label{fig:H_ns_fitted}%
\end{figure}

The numerical solutions show that the typical behavior of the function $H(t)$ during inflation is a sigmoid curve, which can be approximated by using a generalized logistic function $\bar H(t)$ of the form
\begin{equation}\label{H_logistic}
    \bar H(t)= \frac{H_m}{\left(1+e^{k(t-t_0)}\right)^n},
\end{equation}
where $H_m$, $k$, $t_0$ are constants needs to be determined and $n$ is integer. For the numerical solution presented in Fig.(\ref{fig:H_vs_t_numerical}) with model parameters 
\[
\left[\beta,\, \delta, \gamma,\, \Lambda_0 \right] = \left[ 253/(1440\pi^2),\, -0.25,\, 0.0043,\, 0 \right],
\]
the fitted values of the constants in $\bar H(t)$ are $\left[H_m, \, k,\, t_0 \right] = \left[4.389,\, 0.414,\, 35.18\right]$ for the minimum possible integer $n=3$, but in general, a higher $n$ could better fit. We have presented $\bar H(t)$ in Fig.(\ref{fig:H_ns_fitted}-a), and it can be seen that it coincides with numerical solutions during the inflation and decomposed from it when $\epsilon \approx \mathcal{O}(1)$, and oscillations starts. The logistic function $\bar H(t)$ can be used for an approximate analytical solution for the inflationary period; hence, in the next section when we analyze the matter perturbations, this approximate form is used for the expression of $H(t)$.  

%%%%%%%%%%%%%%%%%%%%%%%%%%%%%%%%%%%%%%%%%%%%%%%%%%

\subsection{Evolution of matter perturbations}

The theory deserves the study of cosmological perturbations in full detail; however, since this is rather cumbersome work, let us make a numerical analysis following the work \cite{Laurentis}. Consider the equation which governs the evolution of the matter fluctuations in the linear regime
\begin{equation}\label{matter_fluctuations}
    \ddot{\delta}_{m}+2 H \dot{\delta}_{m}- \frac{1}{2} \kappa_{\mathrm{eff}} \rho_{m} \delta_{m}=0 
\end{equation}
where $\rho_{m}$ is the matter density and $\kappa_{\text {eff }}$ is the effective gravitational coupling constant which, in $\mathcal{F}(R,\mathcal{G})$ gravity case, is

\begin{equation}
    \kappa_{\mathrm{eff}}=\frac{\kappa_0}{\mathcal{F}_{R}(R, \mathcal{G})} 
\end{equation}
where $\kappa_{0}$ is the Newton gravitational constant. Notice that for maximally symmetric background, $\kappa_{\mathrm{eff}}=\kappa$ in \eqref{eff_newton}.  Here, we are considering perfect fluid matter that is added in action \eqref{fRG_action} as minimally coupled. We use EoM \eqref{EoM} with matter density contribution as

\begin{equation}
    \rho_{m}= \frac{3 H^{2}}{\kappa_0} - \rho_{\mathrm{eff}} 
\end{equation}
where $\rho_{\mathrm{eff}}$ includes contributions of both modified gravity, $\rho_{\mathrm{MG}}$, and trace-anomaly, $\rho$ in \eqref{rho_QC}. 

\begin{figure}[H]%
    \captionsetup{format=plain, font=footnotesize, labelfont=bf}
    \centering
    \subfloat[\centering matter perturbations with respect to scale factor ]
    {{\includegraphics[width=9cm]{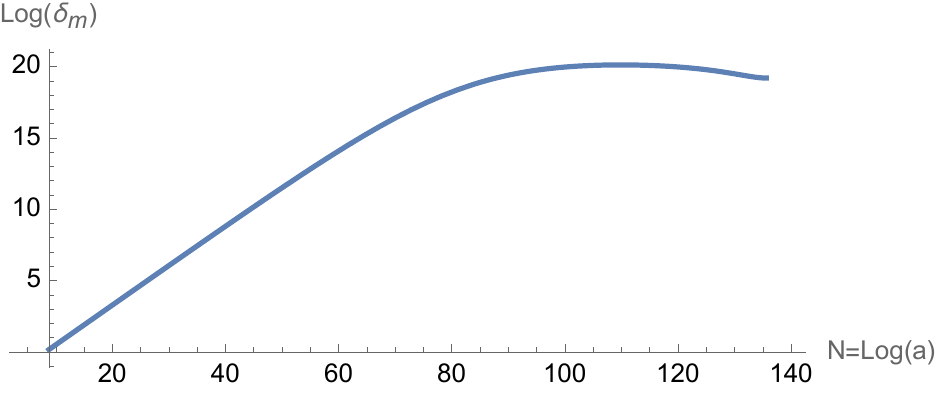} }}%
    \caption{Evolution of the matter perturbations during the inflation in with respect to scale factor. The matter perturbations are linearly grows during the initial phase, and almost frozen at the last stage. The Log-Log plot is in base $e$.}%
    \label{fig:matter_perturb}%
\end{figure}

Substituting these into \eqref{matter_fluctuations}, and by using approximate solution of $H=\bar H(t)$ as a background expansion, one can find a numerical solution for the matter fluctuations $\delta_m(t)$. With the parameters chosen in previous section, the evolution of the matter perturbations during the inflation is given in Fig.(\ref{fig:matter_perturb})

We observe linear growth in matter perturbations concerning the scale factor during the initial phase. Perturbations reach a maximum value, about a factor of $10^9$  in our case, and almost freeze by the end of inflation. This gives an overview of the matter perturbations, but a full cosmological perturbation theory with standard SVT decomposition in the metric should be studied for more accurate results.

%%%%%%%%%%%%%%%%%%%%%%%%%%%%%%%%%%%%%%%%%%%%%%%%%%

\section{Conclusions}

We have studied the cosmological inflation era in a theory that extends General Relativity in such a way that the resultant theory shares some nice properties of the former, such as the uniqueness of the maximally symmetric vacuum and the non-existence of degrees of freedom other than a massless spin-2 graviton while being a better-behaved theory in the ultraviolet regime of gravity. In principle, the theory is of the Born-Infeld type and includes infinitely many powers of curvature in small curvature expansion; but we studied a particular limit of the theory which has judiciously chosen curvature terms up to and including quartic order. The resulting theory becomes a particular $\mathcal{F}(R, \mathcal{G})$ theory where $R$ is the scalar curvature and $\mathcal{G}$ is the Gauss-Bonnet invariant. This theory, when augmented with the inevitable trace anomaly as computed in the Standard Model and the Minimal Supersymmetric Standard Model, allows a quasi-de Sitter solution with enough number of $e$-foldings. To describe the typical behavior of the inflationary dynamics, we studied phase-portrait analysis and also give numerical solutions for $H(t)$. In the Supersymmetric Model, the cubic and quartic terms allow an exit from the otherwise stable de Sitter phase if the initial value of the time derivative of the Hubble parameter is large enough while still in the slow-roll regime. 

%%%%%%%%%%%%%%%%%%%%%%%%%%%%%%%%%%%%%%%%%%%%%%%%%%

\section{Appendix A: Potential analysis of the equivalent point-like Lagrangian} \label{appendix_potential}

As discussed, the $\Box R$ term in the trace of anomalous energy-momentum tensor is equivalent to having $R^2$ term in the Lagrangian, and the renormalization dependence of this term gives us freedom to add the $R^2$ term to the Lagrangian to adjust the value of $\delta$. Here, we investigate how this term in the Lagrangian affects the ``potential term'' of the equivalent point-like Lagrangian of our quartic theory \eqref{quartic_lagrangian}.

\begin{figure}[H]%
    \captionsetup{format=plain, font=footnotesize, labelfont=bf}
    \centering
    \subfloat[\centering without $R^2$ term]
    {{\includegraphics[width=7cm]{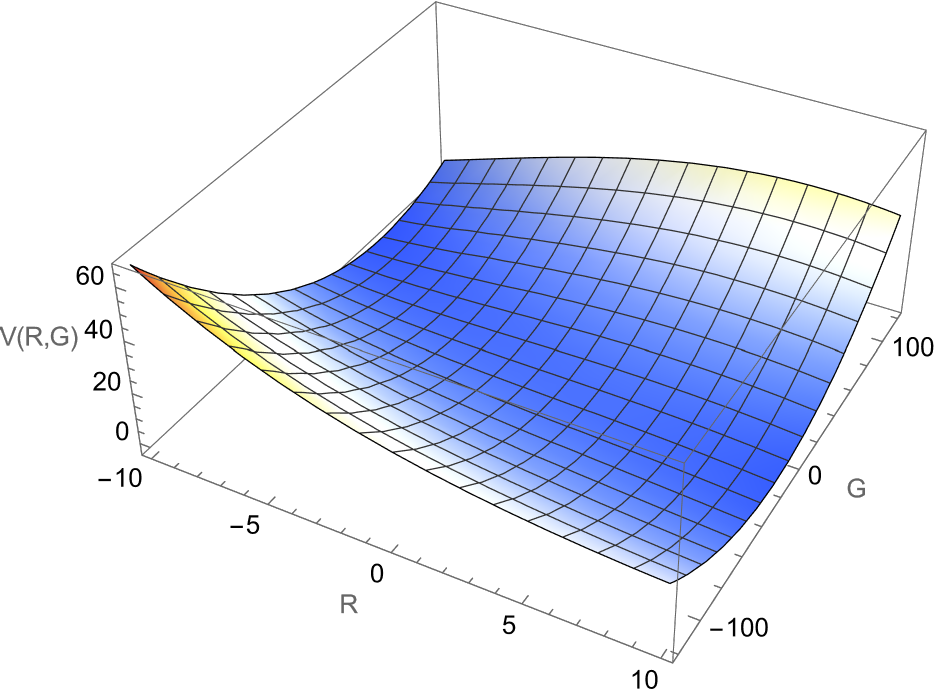} }}%
    \,
    \subfloat[\centering with $R^2$ term]
    {{\includegraphics[width=7cm]{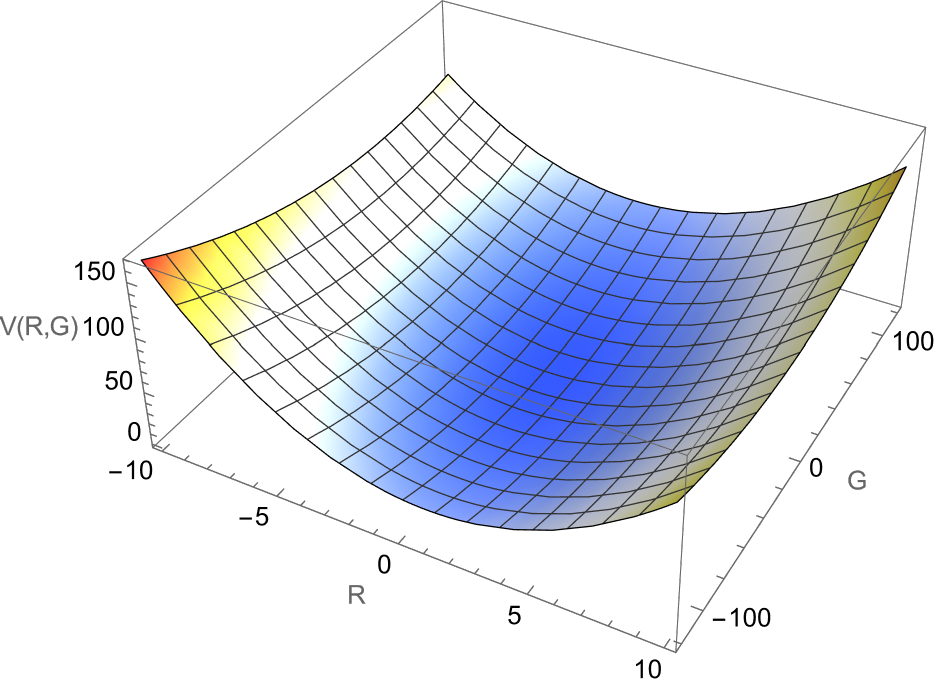} }}%
    \caption{Plot of $V(R,\mathcal{G})$ of the point-like Lagrangian which is equivalent to \eqref{fRG_with_R2}. When the $R^2$-term is dominant (right panel), there is a minimum potential located at $(R,\,\mathcal{G})=(0,0)$ where the dynamical system flows through there.}%
    \label{fig:3D_potentials}%
\end{figure}

\begin{figure}[H]%
    \captionsetup{format=plain, font=footnotesize, labelfont=bf}
    \centering
    \subfloat[\centering without $R^2$ term]
    {{\includegraphics[width=7cm]{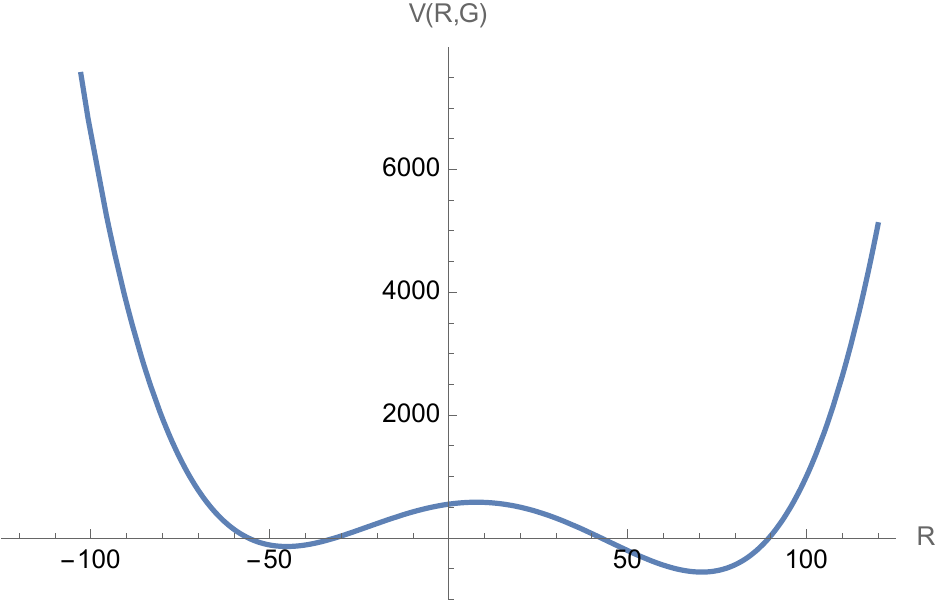} }}%
    \,
    \subfloat[\centering with $R^2$ term ]
    {{\includegraphics[width=7cm]{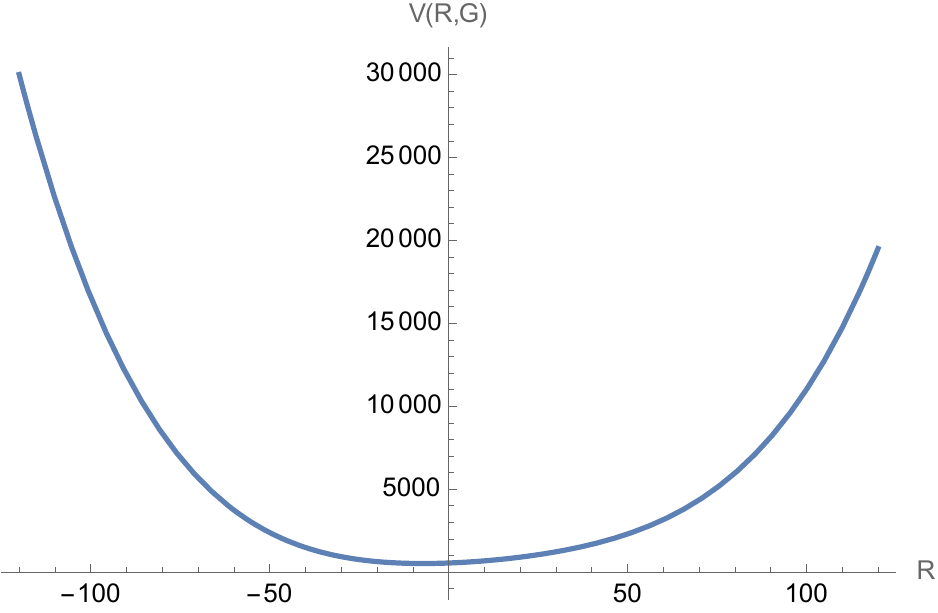} }}%
    \caption{Plots of sections of the potential $V(R, \mathcal{G})$. In the left panel, we reported the section of the potential when the $R^{2}$ term is absent. A symmetry breaking is evident. In the right panel, when $R^2$ term dominant, the minimum of the potential is shifted to $R=0$}%
    \label{fig:2D_potentials}%
\end{figure}

For a generic $\mathcal{F}(R,\mathcal{G})$ Lagrangian, it is showed in \cite{Capozziello2014} that a point-like Lagrangian can be obtained for the metric \eqref{RW}, so inserting it into the generic $\mathcal{F}(R,\mathcal{G})$ action \eqref{fRG_action} and assuming suitable Lagrange multipliers for $R$ and $\mathcal{G}$, we can express the equivalent form as
\begin{equation}
    \mathcal{L}=  6 a \dot{a}^{2} \mathcal{F}_{R}+6 a^{2} \dot{a} \dot{\mathcal{F}}_{R}-8 \dot{a}^{3} \dot{\mathcal{F}}_{\mathcal{G}} +a^{3}\left[\mathcal{F}(R, \mathcal{G})-R \mathcal{F}_{R}-\mathcal{G} \mathcal{F}_{\mathcal{G}}\right]
\end{equation}
which is a canonical function depending on $t$ and defined in the configuration space $\mathcal{Q} \equiv\{a, R, \mathcal{G}\}$, and the Lagrange multipliers are
\begin{equation}
    R=6\left(2 H^{2}+\dot{H}\right)  \,, \qquad
    \mathcal{G}=24 H^{2}\left(H^{2}+\dot{H}\right)  \,.
\end{equation}
Then, this Lagrangian can be decomposed as
\begin{equation}
    L=K\left(q_i, \dot{q}_j\right)-V\left(q_i\right)
\end{equation}
where $K$ and $V$ are the kinetic energy and potential energy respectively. Here we have $q_{i} \equiv\{a, R, \mathcal{G}\}$ and $\dot{q}_{j} \equiv\{\dot{a}, \dot{R}, \dot{\mathcal{G}}\}$. In the case of Lagrangian, considering the Lagrangian density, i.e. $\mathcal{L}=a^{3} L$, they are

\begin{equation}   
    K(a, \dot{a}, R, \dot{R}, \mathcal{G}, \dot{\mathcal{G}})=6\left(\frac{\dot{a}}{a}\right)^{2} \mathcal{F}_{R}+6\left(\frac{\dot{a}}{a}\right) \dot{\mathcal{F}}_{R}-8\left(\frac{\dot{a}}{a}\right)^{3} \dot{\mathcal{F}}_{\mathcal{G}} \,,
\end{equation}
\begin{equation} 
    V(R, \mathcal{G})=-\left[\mathcal{F}(R, \mathcal{G})-R \mathcal{F}_{R}-\mathcal{G} \mathcal{F}_{\mathcal{G}}\right] \,.
\end{equation}
Eventually, for our Lagrangian with $R^2$ term added, 
\begin{equation}\label{fRG_with_R2}
    \mathcal{F} = R-2\Lambda_0 
    + \frac{9 \gamma}{2} {\cal G}
    + \frac{\gamma ^2}{2}  \left(9 {\cal G} R-R^3\right) 
    + \frac{\gamma ^3}{8} \left(81 {\cal G}^2-18 {\cal G} R^2+R^4\right)
    + \kappa_0 \alpha R^2
\end{equation}
we found the potential term as
\begin{equation}
   V(R,\mathcal{G}) = \gamma ^3 \left(\frac{81 \mathcal{G}^2}{8}-\frac{9 \mathcal{G} R^2}{2}+\frac{3
   R^4}{8}\right)+\gamma ^2 \left(\frac{9 \mathcal{G}
   R}{2}-R^3\right)+\alpha  R^2
\end{equation}
We plot the potential with and without the $R^2$ term in Fig.(\ref{fig:3D_potentials}), and in Fig.(\ref{fig:2D_potentials}), we plot a section of the potential for high value of $\mathcal{G}$ (recall that $R\sim H^2$ and $\mathcal{G}\sim R^2$). It seems that without $R^2$ term, the minimum of the potential is not located at $R=0$ and there are two local minimum, one of which is lower on the positive $R$ side. However, adding $R^2$-term changes the potential so that the minimum of the potential is shifted to $\{ R, \mathcal{G} \}\rightarrow 0$. This means that, if the Universe started with some value of $H$, due to $\beta$ term in the trace anomaly for example, then it rolls down to the minimum of the potential which is located at $(R,\,\mathcal{G})=(0,0)$ by inflating the Universe.

%%%%%%%%%%%%%%%%%%%%%%%%%%%%%%%%%%%%%%%%%%%%%%%%%%

\section{Appendix B: Degrees of Freedom in the Quartic Theory}

To figure out the excitations and their masses around a maximally symmetric solution of the minimal quartic theory plus the $\alpha R^2$ term coming from the anomaly in the action, we will follow the discussion given in \cite{Tekin3} for generic $f(\text{Riemann})$ theories.
Let us start with the full field equations in a vacuum:
 \begin{align}\nonumber
        & {\cal F}_R  R_{\mu \nu}+\frac{1}{2}g_{\mu \nu}
        ({\cal {G}} {\cal F}_{\cal G}- {\cal F}) +(g_{\mu \nu}\square -\nabla_\mu \nabla_\nu  )  {\cal F}_R \\
        & +4\left[ (2 C_{\mu \sigma \nu \lambda}-R_{\mu \sigma \nu \lambda} )\nabla^\sigma \nabla^\lambda
        +\frac{R}{6}(g_{\mu \nu}\square -\nabla_\mu \nabla_\nu )\right] {\cal F}_{\cal G}
        = 0,
    \label{full_eqn1}     
\end{align}  
where, for our quartic theory,
\begin{equation}
    \mathcal{F} = R-2\Lambda_0 
    + \frac{9 \gamma}{2} {\cal G}
    + \frac{\gamma ^2}{2}  \left(9 {\cal G} R-R^3\right) 
    + \frac{\gamma ^3}{8} \left(81 {\cal G}^2-18 {\cal G} R^2+R^4\right)
    + \kappa_0 \alpha R^2
\end{equation}
and the relevant partial derivatives
${\cal F} (R,{\cal G})$ read
\begin{eqnarray}
    &&{\cal F}_{\cal G} :=\frac{\partial  {\cal F}}{\partial {\cal G}} = \frac{9}{4}  \gamma \left( -\gamma^2 R^2+9 \gamma^2 {\cal G} +2 \gamma R +2\right), \nonumber \\ 
    &&{\cal F}_R := \frac{\partial  {\cal F}}{\partial R} =\frac{1}{2} (\gamma R-1 ) \Big (\gamma R (\gamma R-2 )-9 \gamma^2 {\cal{G}} -2 \Big ) + 2 \kappa_0 \alpha R.
\end{eqnarray}
Note that the last term comes from the $\alpha R^2$ correction in the action. 
To understand how many independent linear wave equations are encoded in (\ref{full_eqn1}), let us first find its constant curvature vacua and linear the field equations around any of them.
For constant curvature solutions, one has 
\begin{equation}
    \bar{R}_{\mu \sigma \nu \rho}=\frac{\Lambda}{3}(\bar{g}_{\mu \nu}\bar{g}_{ \sigma \rho}-\bar{g}_{\mu \rho}\bar{g}_{\sigma \nu}), \hskip 0.7 cm  \bar{R}_{\mu \nu} = \Lambda \bar{g}_{\mu \nu}, \hskip 0.7 cm \bar R = 4 \Lambda, 
\end{equation}
$\Lambda$ being the effective cosmological constant to be determined now. For the maximally symmetric solution (\ref{full_eqn1}) reduces to 
\begin{equation}
    \bar{R}\,\bar{\cal F}_R+2 \bar{{\cal {G}}}\,\bar{\cal F}_{\cal {G}} -2\bar{\cal F}=0,
    \label{bartrace1}               
\end{equation} 
In four dimensions, the $\alpha R^2$ term does not contribute to the maximally symmetric solution, so one has 
\begin{equation}
    4 \lambda^4 + 4 \lambda^3 - \lambda + \lambda_0=0,
    \label{vac_denk2}
\end{equation}
where $\lambda = \gamma \Lambda$ as was used in the bulk of the paper. We do not need the explicit solutions of this equation, the fact that at least one viable solution is sufficient for us. There exists, in fact, only one viable solution, we do not want to repeat this discussion here, see \cite{esin}. Let us now linearize the field equations about this solution. Let 
\begin{equation}
    g_{\mu \nu}= \bar{g}_{\mu \nu}+h_{\mu \nu}, 
\end{equation}      
then one can write the linearization of the Gauss-Bonnet scalar ${\cal {G}}_L$   in terms of linearized scalar curvature $R_L$ as  
\begin{equation}
    {\cal {G}}_L  =\frac{4}{3}\Lambda R_L,
\end{equation} 
where $\Lambda$ is the viable solution of (\ref{vac_denk2}). In terms of the perturbation $h_{\mu \nu}$, one has 

\begin{equation}
    R_L = - (\bar{\square} + \Lambda) h + 
    \bar{\nabla}^\mu \bar{\nabla}^\nu h _{\mu \nu},
\end{equation}
where $h \equiv \bar{g}^{\mu \nu} h_{\mu \nu}$ and barred quantities refer to the background metric.  One has  
\begin{eqnarray}
    ( {\cal F}_R)_L  &=& \left (-6  \gamma \lambda (1+2 \lambda )+2 \kappa_0 \alpha \right) R_L, \hskip 0.5 cm  
    ( {\cal F}_{\cal G})_L  = \frac{9}{2} \gamma^2 (1+2 \lambda )R_L, \nonumber \\
    ( {\cal F})_L  &=&  \left((1+2 \lambda )^3 + 8 \kappa_0 \alpha \Lambda \right)R_L. 
    \label{linearized_quant}    
\end{eqnarray}
Linearization of (\ref{full_eqn1}) around the background solution yields
\begin{align}\nonumber
    & ({\cal F}_R)_L \Lambda \bar g_{\mu \nu}+\bar {\cal F}_R (R_{\mu \nu})_L +
    \frac{1}{2}h_{\mu \nu}
    (\bar {\cal {G}} \bar {\cal F}_{\cal G}- \bar {\cal F}) + \frac{1}{2}\bar g_{\mu \nu}
    \Big (({\cal {G}})_L \bar {\cal F}+\bar {\cal {G}} ({\cal F}_{\cal G})_L- ({\cal F})_L \Big) \nonumber \\
    &  +(\bar g_{\mu \nu}\bar \square -\bar \nabla_\mu \bar \nabla_\nu  ) \Big ( ({\cal F}_R)_L +\frac{4 \Lambda}{3}({\cal F}_{\cal G})_L \Big)
    = 0,
    \label{full_eqn_lin}     
\end{align}  
 where $ \bar {\cal G}= \frac{ 8 \Lambda^2}{3}$. Defining the linearized version of the cosmological Einstein tensor $G_{\mu \nu }= R_{\mu \nu} - \frac{1}{2} g_{\mu \nu} R + \Lambda g_{\mu \nu}$ as $(G_{\mu \nu})_L$, one can rewrite (\ref{full_eqn_lin} ) as
\begin{equation}\nonumber
    \left (8 \alpha \kappa_0 \Lambda + \frac{\kappa_0}{\kappa}\right)  G_{\mu \nu }^L
    +2 \alpha \kappa_0 (\bar g_{\mu \nu}\bar \square -\bar \nabla_\mu \bar \nabla_\nu +\Lambda g_{\mu \nu} )R_L
    = 0,
    \label{full_eqn_lin2}     
\end{equation}  
where we defined
\begin{equation}
    \frac{1}{\kappa} = \frac{1}{\kappa_0} 
    (1+ 2 \lambda)^2(1- 4 \lambda)
\end{equation}
Without the $R_L$ part in (\ref{full_eqn_lin2}), we already know that the theory is cosmological Einstein's gravity and only excitation defined by $G_{\mu \nu }^L=0$ is the massless spin-2 equation. But with $\alpha \ne 0$, the equation (\ref{full_eqn_lin2}) describes a coupled system of massless spin-2 and massive spin-0 particles. We refer the reader to \cite{Tekin3} for decoupling these modes. Here let us calculate the mass of the spin-0 excitation. For this, taking the trace of   (\ref{full_eqn_lin2}) yields 
the massive Klein-Gordon equation in de Sitter spacetime: 
\begin{equation}
    \left (\bar \square -\frac{1}{6 \kappa \alpha} \right) R_L =0.
\end{equation}
Then one has the mass of the scar field as  $m_s^2 = \frac{1}{6 \kappa \alpha}$. Note that if one adds a $\beta R_{\mu \nu}R^{\mu \nu}$ term to the action, the spectrum includes a massive spin-2 particle which is necessarily ghost-like due to its conflict with the massless spin-2 particle \cite{ Stelle,Tekin3}.

%%%%%%%%%%%%%%%%%%%%%%%%%%%%%%%%%%%%%%%%%%%%%%%%%%

\section*{Acknowledgment}  We would like to thank Prof. Dr. Sakir Erkoc for their kind support.

\end{document}